\documentclass[12pt, times]{article}

\usepackage{times}
\usepackage{fullpage}
\usepackage{graphicx}
\usepackage{amsmath}
\usepackage{amssymb}
\usepackage[numbers,sort]{natbib}

\usepackage[hyphens]{url}

\usepackage[
bookmarksnumbered = true,
bookmarksopen = true,
bookmarksopenlevel=3,
pdftitle={Simple and Robust Boolean Operations for Triangulated Surfaces},
pdfauthor={Gang Mei},
pdfstartview = FitH,
colorlinks=true,
linkcolor=blue,
breaklinks=true,
urlcolor=black,
citecolor=blue
]{hyperref}

\usepackage{multicol}        
\usepackage{float}
\usepackage{multirow}
\usepackage{algorithm} 
\usepackage{algorithmic} 

\usepackage{booktabs} 
\usepackage{subfigure}

\usepackage{fontenc}
\usepackage{enumerate}
\usepackage[perpage,para,symbol*]{footmisc}

\usepackage{listings} 
\usepackage{xcolor} 
\usepackage[letterpaper,twoside,vscale=.8,hscale=.75,nomarginpar,hmarginratio=1:1]{geometry}

\title{Simple and Robust Boolean Operations for Triangulated Surfaces}
\author{ Gang Mei and John C. Tipper\\
\normalsize         Institute of Earth and Environmental Science, University of Freiburg \\
\normalsize         Albertstra\ss{}e 23B, D-79104, Freiburg im Breisgau, Germany \\
\normalsize         \href{mailto:gang.mei@geologie.uni-freiburg.de}{\{gang.mei, john.tipper\}@geologie.uni-freiburg.de}
}
\date{}

\begin{document}
\maketitle

\begin{abstract}
Boolean operations of geometric models is an essential issue in computational geometry. In this paper, we develop a simple and robust approach to perform Boolean operations on closed and open triangulated surfaces. Our method mainly has two stages: (1) We firstly find out candidate intersected-triangles pairs based on Octree and then compute the intersection lines for all pairs of triangles with parallel algorithm; (2) We form closed or open intersection-loops, sub-surfaces and sub-blocks quite robustly only according to the cleared and updated topology of meshes while without coordinate computations for geometric entities. A novel technique instead of inside/outside classification is also proposed to distinguish the resulting union, subtraction and intersection. Several examples have been given to illustrate the effectiveness of our approach.

\end{abstract}

\section{Introduction}
\label{sec:1}
The calculating of Boolean operations for solids plays an important role in geometric processing, which intends to obtain union, subtraction and intersection of geometric models. The most commonly used method to display geometric models probably is the boundary representation (B-Rep) \cite{Hoffmann1989} which can be based on parametric surfaces such as NURBS surfaces or discrete surfaces. The discrete surface is the decomposition of a specific surface domain, of which polygonal meshes especially triangular meshes is the most popular kind of representation form.

In this paper, we focus our contribution on the implementation of Boolean operations over triangular meshes. As mentioned above, triangular meshes have become the most popular surface representation. They are normally required to be manifold in most applications, however this property may be lost in some cases such as undergoing dynamical collision or large deformation. Thus reconstruction need to be carried out over those meshes to keep them manifold \cite{Zaharescu2011}. In this paper, we perform simple and robust Boolean operations on a pair of manifold triangulated surfaces without considering the case of self-intersecting meshes.

In our algorithm, surfaces are classified into two types: open or closed. Open surfaces have outer or inner boundary loops while closed ones without. When a surface is closed, the geometric model bounded by it can be called as a volume or a block. The geometric objects we will manipulate are a pair of manifold triangulated surfaces which can be open-and-open, open-and-closed or closed-and-closed.

\subsection{Related works}
\label{sec:1.1}
Varies of implementations of Boolean operations are described in the literatures. They can be roughly classified according to three main properties: the type of input data, the type of computation and the type of output data \cite{HybridBooleans2010}. About the type of computation, Tayebi et al. \cite{Tayebi2011} assort them into four different categories: exact  arithmetic methods \cite{ExactRobust2010}, approximate arithmetic methods \cite{Smith2007}, interval computation methods and volumetric methods \cite{Chen2007, wang2011}.

Most of approaches described in the literatures specially aim at one type of the input models such as those by B-Rep based on NURBS surface \cite{Tayebi2011, yang2009}, curved surface \cite{LoSH2005} and polygonal meshes \cite{ExactRobust2010, HybridBooleans2010, Schifko2010}. Besides these popular surface-based representations, several other methods like Nef \cite{Nef2007}, surfel-bounded \cite{Adams2003}, L-Rep \cite{L-Rep2010} for representing geometric models are also developed to construct new models or convert existing ones to target models, and then Boolean operations are also implemented. In another aspect, the input models can be manifold or non-manifold \cite{Pereira2011}, and most methods perform Booleans on manifold ones.

Among the types of computation, the exact  arithmetic methods and interval computation methods need directly computing Booleans over the initial surfaces, while the rest of two do not and hence deem to be indirect.

When implementing direct Booleans for mesh models, there are two key procedures which strongly affect the effectiveness and efficiency of the whole algorithm. The first one is how to robustly obtain the intersection lines and loops of all intersected entities as faster as possible. The core of such procedure is to accurately find out all potentially intersected entities in a short time to reduce computation cost. Many techniques based on BSP(binary space partitions) \cite{ExactRobust2010}, Octree \cite{HybridBooleans2010}, OBB trees \cite{OBBTrees2009}, bipartite graph structure \cite{Severn2006} and tracking neighbors \cite{STL2007, LoSH2005} have been developed to realize this goal. The second key procedure is how to correctly assemble and distinguish the union, subtraction and intersection of intersected models. The most direct method is to make a inside/outside classification \cite{Chen2010}, which checks the location of vertices or facets with the resulting volumes. Varies of algorithms of direct Boolean operations develop their own solutions for the above two problems.

Quite recently, several novel methods of Boolean operations on polygonal meshes have been presented. Pavi\'{c} et al. \cite{HybridBooleans2010} develop a hybrid method which combines polygonal and volumetric computations and representations for performing Boolean operations over polygonal meshes; similarly Garc\'{i}a et al. \cite{Garcia2011} use Extended Simplicial Chains(ESCs) to represent both boundary and volume of free-form solids when perform Booleans. Aiming at robustness of Boolean operations for large number of triangles in industrial applications, Schifko et al. \cite{Schifko2010} adopt several suitable libraries from CGAL and GNU multi precision arithmetic library to filter exact arithmetic. An approximate method based on Layered Depth Images(LDI) for polygonal models is developed in  \cite{wang2011}, in which LDI is accepted to sample and trimmed adaptive contouring is used to rebuild intersected meshes.Campen and Kobbelt \cite{ExactRobust2010} adopt plane-based representations, BSP techniques and a localization scheme to obtain  exact and robust (self-) intersections for polygonal meshes.

Besides the well-known libraries CGAL \cite{CGAL2012}, several open source packages (MeshLab \cite{MeshLab2008}, OpenFlipper \cite{OpenFlipper2010}, MEPP \cite{MEPP2012}) also contains robust implementations of Boolean operations.

\subsection{Our contribution}
\label{sec:1.2}
When designing algorithms to solve a specific problem, normally a fast algorithm with better efficiency also has higher complexity than a slow one. This is certainly true for performing Boolean operations on triangular meshes. In this paper, we will try to keep a relative balance between the efficiency and complexity of such algorithms. And our objective is to develop a simple and robust approach of Boolean operations which also have satisfied efficiency. Several highlights can be drawn in our method:

(1) A simple and relative fast method to calculate the intersection lines of triangles.
In order to reduce computation cost, we firstly use the spatial decomposition data structure Octree to locate and search candidate intersected triangle-pairs, and then compute the intersection line of each pair of triangles parallel.

(2) A fast approach only based on operations of the indexes of entities and without computation for coordinates to obtain intersection loops, sub-surfaces and sub-blocks.

(3) A novel technique to distinguish the union, intersection and subtraction volume of two intersected closed surfaces(solid models). This simple classification method is very fast and robust since it is also only based on operations of the indexes of entities.

The rest of this paper is organized as follows. In Sect.2, we give a brief view to our algorithm; in Sect.3 we design some data structure and define several notations for our implementation; and then, all details of the approach will be described in Sect.4; Then, we give several examples to illustrate the effectiveness of the approach in Sect.5, and finally conclude and discuss our work in Sect.6.

\section{Overview of the approach}
\label{sec:2}

\begin{figure}[htbp]
\centering
\includegraphics[height = 75mm]{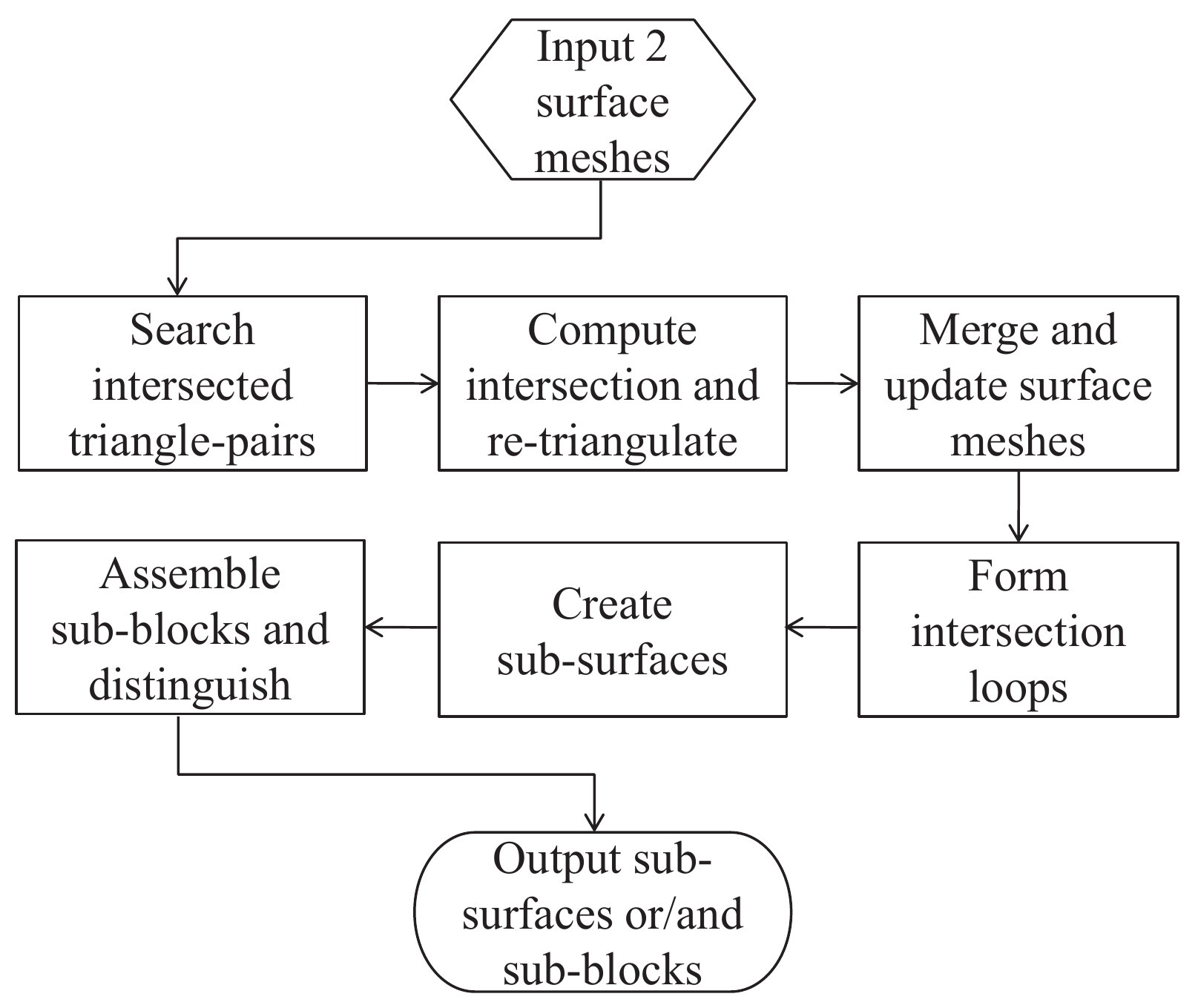}
\caption{Flow of the approach}
\label{fig:1:Flow}       
\end{figure}

Our method can be mainly summarized into 6 steps (Fig.\ref{fig:1:Flow}), and can be divided into two states: the first, which is based on computation for coordinates of entities, includes searching intersected triangle-pairs, calculating intersection line of each pair of triangles, re-triangulation and updating the resulting surface meshes; the second  procedure is only based on operations for indexes of entities including forming intersection loops, creating sub-surfaces, assembling and distinguishing sub-blocks.

Step 1: Search the candidate intersected triangle-pairs. In order to reduce the cost on computing, a robust and fast search algorithm is needed to obtain potentially intersected triangles. In this paper, we use Octree \cite{RealTimeCD2005} to locate and find out all candidate intersected triangle-pairs.

Step 2: Compute the intersection line for each pair of triangles. M\"{o}ller \cite{Moller1997} developed a robust and efficient algorithm to do this and we adopt his work. When there are a huge number of triangle-pairs needed to calculate their intersections, parallel computation based on OpenMP \cite{OpenMP}  is accepted.

Step 3: Merge and renumber all vertices, update and clear meshes. New vertices will be produced after intersecting of triangle-pairs, and the intersected triangles are replaced by re-triangulation. In order to keep a valid topology, all vertices are merged and renumbered and all triangles are updated.

Step 4: Connect intersection lines into closed or open loops. After computing the intersection of each pairs of triangles, a set of discrete edges can be obtained, and they need to be connected into closed or open orientated loops. If there not exists at least one closed loop on an intersected surface, then no closed block bounded by triangular facets will be formed.

Step 5: Obtain sub-surfaces based on closed loops. A sub-surface includes the closed loop and all of its incident triangles. The edges of a closed loop are set as the advancing front; "grow" a new surface based on the topology until the number of faces in the sub-surface not increases (Fig.\ref{fig:6b}).

Step 6: Assemble and distinguish sub-blocks. Sub-blocks can be easily created by assembling related sub-surfaces (Fig.\ref{fig:6c}). In further, the boundary closed loops generated in Step 4 of sub-surface can be used to represent the sub-surface. Hence, assembling and distinguishing can be done based on the boundary closed loops.

\section{Data structure and notation}
\label{sec:3}
In this section, some notations for geometric entities are defined. Additional properties such as direction are added onto some of them.

\textbf{Definition 1} \emph{Directed edge} is a directed segment, in which the first vertex is named as Head, while the other is Tail.

\textbf{Definition 2} \emph{Orientated loop} is a set of connected directed edges, which can be closed or open. It can be also represented by a set of ordered vertices. A pair of orientated loops are defined as \emph{twins} if they have same vertices and opposite order.

\textbf{Definition 3} \emph{Normalized face} is a coplanar polygon with its normal, which is triangle or polygon in this paper.

To describe our algorithm in following sections, we notate several common geometric objects and allocate arrays to store geometric entities.

\textbf{m\_aVerts:} an array of vertex to store all vertices after intersection of all triangle-pairs  and merging; all vertices must be checked and renumbered.

\textbf{m\_aEdges:} an array of edge to store all intersection lines after intersecting of all triangle-pairs; the Head and Tail of each edge in this array are updated after merging and renumbering all vertices.

\textbf{m\_aLoops:} an array of loop to store all closed/open intersection loops. The set of ordered vertices in each loop must have the newly updated IDs.

\textbf{m\_aPolys:} an array of polygon to store all resulting polygons produced from the intersecting of triangle-pairs. Each polygon in this array will be decomposed into new triangles via polygon triangulation.

\textbf{m\_aTrgls:} an array of triangle to store all updated triangles after intersecting and merging; the triangles come from (1) the original triangles that do not intersect with others, or (2) the new triangles generated by re-triangulating the resulting polygons from the intersecting of triangle-pairs.

\textbf{m\_aSurfs:} an array of surface to store all created sub-surfaces. Each sub-surface has its boundary sub-loop(s) preparing for assembling sub-blocks.

\textbf{m\_aBlocks:} an array of block to store all assembled sub-blocks.

\section{The Methodology}
\label{sec:4}
In this section, we will present the detailed description to all six steps in our approach separately. Several procedures such as clearing triangular mesh, creating sub-surfaces and sub-blocks will be explained with pseudo codes.

\subsection{Searching intersected triangle-pairs}
\label{sec:4.1}
Before calculating the intersection of any pair of triangles, we should know which pair of triangles potentially intersect. The direct and inefficient method is to make a bounding box intersection test for each triangle in a surface with that of another one. In order to improve the efficiency, we adopt Octree to locate and then find out candidate intersected triangle-pairs.

Given two surface meshes, S$_{A}$ and S$_B$, compute their smallest AABBs denoted as Box$_A$ and Box$_B$, respectively; and then calculate the intersection Box$_{AB}$ of Box$_A$ and Box$_B$(Box$_{AB}$ = Box$_A$$\cap$Box$_B$); check each triangle of S$_A$ and S$_B$ whether it is outside of the volume Box$_{AB}$, and divide S$_A$ and S$_B$ into two sub-arrays where S$_{Aout}$ + S$_{Ain}$ = S$_A$, S$_{Bout}$ + S$_{Bin}$ = S$_B$; and then extend the volume Box$_{AB}$ into a cube to be an AABB for the triangles from both S$_{Ain}$ and S$_{Bin}$ (S$_{Ain}$$\cup$S$_{Bin}$).

Let the bounding cube as the root node of an Octree, and create eight octants for each node recursively. Let the N$_a$ and N$_b$ denote the number of triangles that intersect each interior node from S$_{Ain}$ and S$_{Bin}$ respectively, the recursion of creating 8 octants for each node terminates and then the node becomes a leaf when

 (1)The depth of the node reaches a user-specified maximum depth;

 (2)Both N$_a$ and N$_b$ less than a given allowable number;

 (3)N$_a$ or N$_b$ is zero.

When to check whether a triangle from either S$_{Ain}$ or S$_{Bin}$ is inside a node of the Octree, a simple method is to make an intersection test between the bounding box of the triangle and the node of Octree. if they intersect, the triangle can be considered being inside the node. Noticeably, a same triangle can intersect with several nodes. To reduce the cost on computing bounding box for each triangle, we previously calculate the boxes for all triangles from both S$_{Ain}$ and S$_{Bin}$ once, and then adopt the boxes when need.

\subsection{Intersecting of triangles and re-triangulating}
\label{sec:4.2}
The intersecting of a pair of triangles is not complex, M\"oller \cite{Moller1997} has developed a robust and efficient algorithm and its code to do this. In this paper, we directly adopt the work from M\"oller.

Our interest is that, when there are still a huge number of pairs of triangles needed to calculate their intersections, parallel algorithm is meaningful and necessary here. Computing the intersection of one pair of triangles does not affect the procedure of that to another pair of triangles. The intersecting for any two pairs of triangles can be carried out at the same time. This parallelization can be realized easily by calling the OpenMP API \cite{OpenMP} and adding several very simple routines into the common serial codes. We use the following code structure to compute the intersections of triangle-pairs parallel.

\#pragma omp parallel for \\
\hspace*{1em} for each pair of triangles $p_i$ \{ \\
\hspace*{2em}   Calculate the intersection of $p_i$; \\
\hspace*{2em}   Save the intersection edge of $p_i$ if it exists; \\
\hspace*{1em} \}

\begin{figure}[H]
\centering
\includegraphics[width = 16cm]{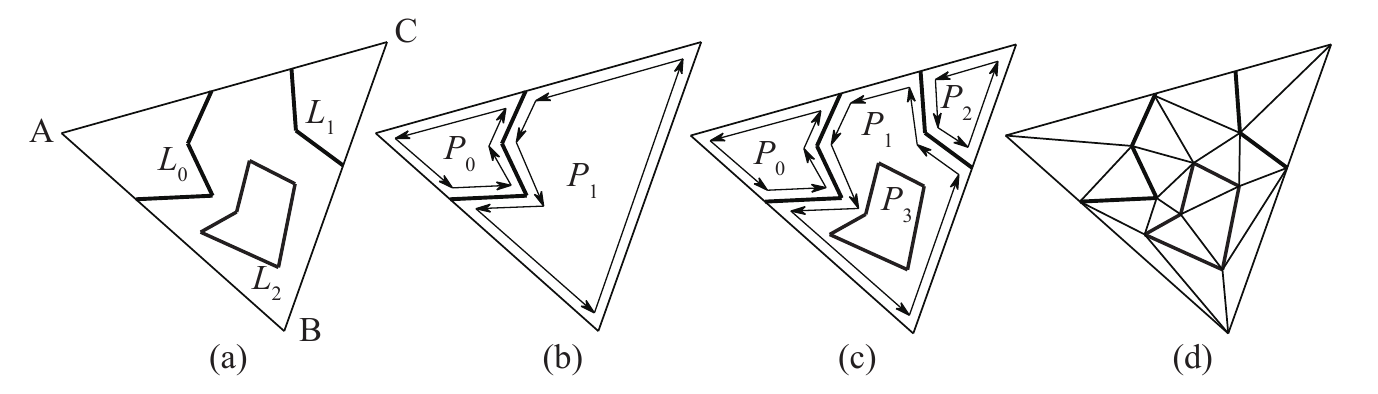}
\caption{Intersected triangle and the re-triangulation. (a) A triangle and its edge-loops; (b) Original triangle divided into 2 polygons; (c) Original triangle divided completely; (d) Triangular partition of the sub polygons.}
\label{fig:3:Re-triangulation}       
\end{figure}

After calculating the intersection for all triangle-pairs, for an individual intersected triangle, usually there are several intersection edges inside the triangle since it perhaps intersects with several other triangles, and it will be divided into several sub polygons by the intersection edges. To generate these sub polygonal faces for a triangle $T_i$, we firstly find all of its intersection edges, and then connect the edges into one or more open or closed loops , and finally divide the newly produced polygons by a unused intersection loop iteratively until all loops of the triangle $T_i$ are adopt, as shown in Fig.\ref{fig:3:Re-triangulation}.

As described above, intersected triangles will be divided into polygonal faces. In order to manipulate the surface mesh more easily in next steps, all resulting polygonal face should be decomposed into triangles via ear-clipping algorithm \cite{EarCut2001}. Since that the ear-clipping is valid for planar and count-clockwise polygons, pre-process must be carried out before re-triangulating a polygon in 3D. For a polygon $P$, its partition can be obtained in 3 steps:

Step 1: Calculate the local coordinate system of $P$, and then transfer $P$ into its local version $P'$;

Step 2: Check whether the vertices of $P'$ orders in count-clockwise (CCW) or clockwise (CW). If CW, reverse $P'$ to make it be CCW;

Step 3: Generate the polygon triangulation $T'$ of $P'$ via ear-clipping, and then directly obtain $T$ for $P$ according to $P'$ for that the topologies of $T$ and $T'$ are exactly the same while the coordinates of vertices differ(Fig.\ref{fig:3:Re-triangulation}(d)).

\subsection{Merging and updating}
\label{sec:4.3}

After intersecting of triangle-pairs, new vertices will be produced and the original intersected triangles are replaced by re-triangulated triangles. In order to get ready for next operations that based on valid topology, the surfaces must be merged and updated. It needs to carry out the following clearings:

(1) merge all vertices and renumber them;

(2) update the indexes of the vertices for each triangle and loop;

(3) check each triangle and loop whether there are same indexed vertices;

\begin{figure}[H]
\centering
\includegraphics[width = 11.8cm]{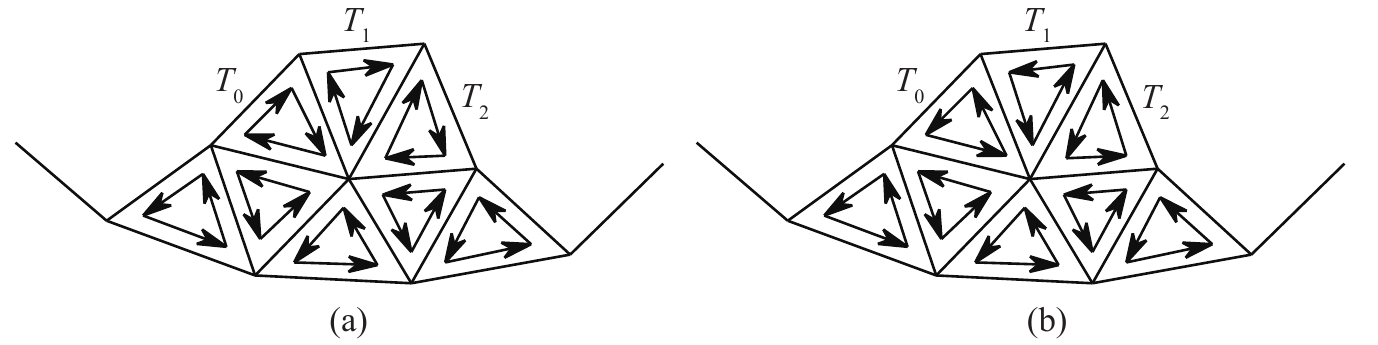}
\caption{ Clear the topology of triangular meshes. (a) Original meshes with invalid triangles $T_0$, $T_1$ and $T_2$; (b) Cleared meshes without same edges}
\label{fig:4:ClearTopology}       
\end{figure}

(4) reverse some newly produced triangles (Fig.\ref{fig:4:ClearTopology}).

Several requirements must be conformed after clearing: no same vertices, no degenerated triangles, no same vertices in a loop and no same edges.

For that the next procedures such as connecting loops and creating sub-surfaces are all based on a valid topology of meshes; thus it is necessary to clear the topology besides merging and renumbering vertices. As shown in Fig.\ref{fig:4:ClearTopology}, (a) is the original triangular meshes with 3 unallowable faces for that they have same edges with their adjacent triangles and (b) is the cleared version.

\subsection{Forming intersection loops}
\label{sec:4.4}
 As defined in Sect.\ref{sec:3}, \textit{Orientated loop} is a set of connected directed edges, which can be \textit{closed} as a cycle or \textit{open} as a polyline. Open loop is the intersection loop in which either the first or the last vertex is only shared by one intersection edge while each of the rest is shared by two edges. Close loop is the intersection loop in which each vertex is shared by at least two intersection edges. we classify the closed intersection loops into: \textit{hard closed} and \textit{soft closed}:

\textit{Hard closed} loop is the one in which each vertex is shared by only two intersection edges; for example, there are six hard closed loops in Fig.\ref{fig:6:CubeSpere}.

\textit{Soft closed} loop is the one in which the first and last vertices are shared by more than two intersection edges while each of the rest is shared by two edges, see 4 soft closed loop in Fig.\ref{fig:12:Cylinder}.

Only closed intersection loops(hard closed and soft closed ones) can be used to create sub-surfaces (Sect.\ref{sec:4.5}). The closed or open loops can be assembled by strictly connecting the Head of an edge with the Tail of another edge, or the Tail of an edge with the Head with another edge. The iteration of assembling will stop when no more connected edges can be found. Besides being displayed by a set of edges, the resulting closed/open loops can be also represented by a set of ordered vertices. For an original surface, there may be several loops on it; and we should know whether a loop is closed or open simply by comparing the Head of the first edge with the Tail of the last edge.

\subsection{Creating sub-surfaces}
\label{sec:4.5}
Assuming there is at least one closed loops in an original surface, then sub-surfaces can be created beginning from the closed loop. A sub-surface includes the closed loop and its incident triangles. Set the edges of a closed loop as the advancing edge front, "growing" until the number of faces in the sub-surface not increases. The growing can be done according the topology of the updated surface. Finding the next face is to search the incident faces of an advancing front edge. After a growing step, all advancing front must be updated to prepare for next growing until a sub-surface is formed; when all closed loops are used, the growing can terminate(Algorithm \ref{alg:3:CreateSubSurface}).

\begin{algorithm}[htb]
\renewcommand{\algorithmicrequire}{\textit{Input:}}
\renewcommand\algorithmicensure {\textit{Output:}}
\caption{Create Sub-surfaces From Closed Loops}
\label{alg:3:CreateSubSurface}
\begin{algorithmic}[1]
\REQUIRE 
A set of closed loops stored in m\_aLoops and a triangular surface S
\ENSURE 
A set of sub-surfaces stored in m\_aSurfs
\FOR {loop $L_i$ in m\_aLoops}
   \STATE create a new empty surface $newSf$;
   \STATE copy and add $L_i$ as a boundary loop into $newSf$;
      \WHILE {the number of triangles of $newSf$ increases}
         \FOR {edge $e_j$ of loop $L_i$}
          \STATE let $egHead$ and $egTail$ be the first and second vertices of $e_j$;
           \FOR {triangle $T_k$ in m\_aTrgls}
              \STATE let nID[3] denote the 3 vertices of $T_k$;
            \IF{(nID[1] = $egHead$ and nID[0] = $egTail$) or
                (nID[2] = $egHead$ and nID[1] = $egTail$) or
                (nID[0] = $egHead$ and nID[2] = $egTail$) }
              \STATE add 3 edges of $T_k$ into loop $L_i$ as new advancing front;
              \STATE add triangle $T_k$ into $newSf$;  break;
            \ENDIF
          \ENDFOR
        \STATE update loop $L_i$ by removing any pair of opposite edges;
       \ENDFOR
     \ENDWHILE
    \STATE add $newSf$ into m\_aSurfs;
\ENDFOR
\end{algorithmic}
\end{algorithm}

\newpage
Generally, there are more than one sub-surfaces can be formed in a original surface, and we classify them into \textit{public} and \textit{private}.

\textbf{Definition 4} \textit{Private sub-surface:} a sub-surface is private when there is only one closed loops in it. ¡°Private¡± means it is owned by a loop privately.

\textbf{Definition 5} \textit{Public sub-surface:} a sub-surface is public when there is more than one closed loops in it. "Public" means several closed loops share this sub-surface together. Noticeably, if this public sub-surface is generated from an original open surface, then it also must have a closed boundary loop besides the intersection closed loop(s).

\textbf{Definition 6} \textit{Sub-surface owner:} the only closed loop which owns the private sub-surface or several closed loops which share the public sub-surface.

\textbf{Remark 1} There is no more than one public sub-surface.

\textbf{Remark 2} There is at least one private sub-surface.

\subsection{Assembling and distinguishing sub-blocks}
\label{sec:4.6}

\subsubsection{Assembling all possible sub-blocks}
\label{sec:4.6.1}

Given two original surface S$_A$ and S$_B$, a set of sub-surfaces will be obtained after intersecting. For a sub-surface $SS$ from the original surface S$_A$, assuming it has n closed loops, noted as $L_i$ ($i = 0,\ldots,n-1$). It also means that there are $n$ owners share the sub-surface S$S$. If we can find $n$ private surfaces from the original surface S$_B$, and each private sub-surface is only owned by the closed loop $L_i$, then a sub-block can be easily assembled by the sub-surface $SS$ from the surface S$_A$ and $n$ private surfaces from S$_B$. Details of assembling sub-blocks is listed in Algorithm \ref{alg:4:CreateSubBlock}. And we can draw some conclusions for assembling sub-blocks.

\begin{algorithm}[htb]
\renewcommand{\algorithmicrequire}{\textit{Input:}}
\renewcommand\algorithmicensure {\textit{Output:}}
\caption{Create Sub-blocks From Sub-surfaces}
\label{alg:4:CreateSubBlock}
\begin{algorithmic}[1]
\REQUIRE 
A set of sub-surfaces stored in m\_aSurfs
\ENSURE 
A set of sub-blocks stored in m\_aBlocks
\WHILE {all sub-surfaces in m\_aSurfs are not tested completely}
    \STATE find a untested sub-surface $startSS$ as the starting one;
    \STATE set $startSS$ as being tested;
    \IF {$startSS$ is a public sub-surface with boundary loop(s)}
    \STATE continue;
    \ENDIF
    \STATE create a new empty sub-block $newBlk$;
    \STATE   add $startSS$ into $newBlk$;
\FOR { the $i^{th}$ closed loop $L_i$ in $startSS$}
\FOR {each sub-surface $SS_i$ in m\_aSurfs }
 \IF { ($SS_i$ is untested and also owned only by $L_i$) and ($SS_i$ and $startSS$ not come from same surface) }
   \STATE set $SS_i$ as being tested;
   \STATE add $SS_i$ into $newBlk$;  break;
   \ENDIF
\ENDFOR
\ENDFOR
\STATE add $newBlk$ into m\_aBlocks;
\ENDWHILE
\end{algorithmic}
\end{algorithm}

\textbf{Remark 3} A private sub-surface can be used once or twice.

\textbf{Remark 4} A public sub-surface without boundary loop can be used once or twice.

\textbf{Remark 5} A public sub-surface with a boundary loop, which is generated from an open original surface, cannot be adopted to assemble sub-blocks. For that it is unable to find a sub-surface owned also by the boundary closed loop in its intersected original surface.

\subsubsection{Distinguishing}
\label{sec:4.6.2}
\begin{figure}[h]
\centering
\subfigure[Original triangular blocks]{
  \label{fig:6a}       
  \includegraphics[width=11.8cm]{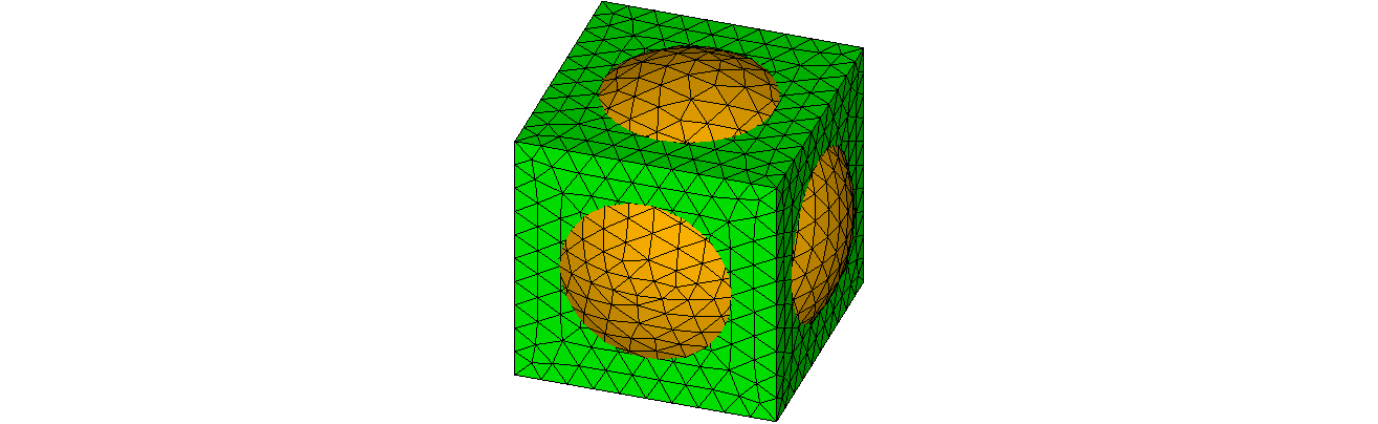} }
\subfigure[Sub-surfaces]{
  \label{fig:6b}       
  \includegraphics[width=11.8cm]{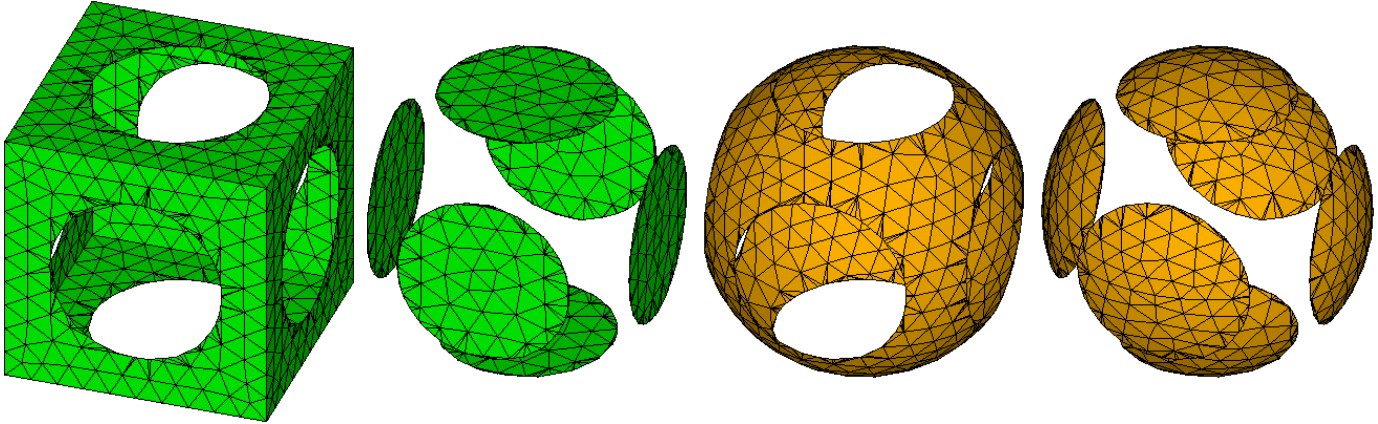} }
\subfigure[Sub-blocks]{
  \label{fig:6c}       
  \includegraphics[width=11.8cm]{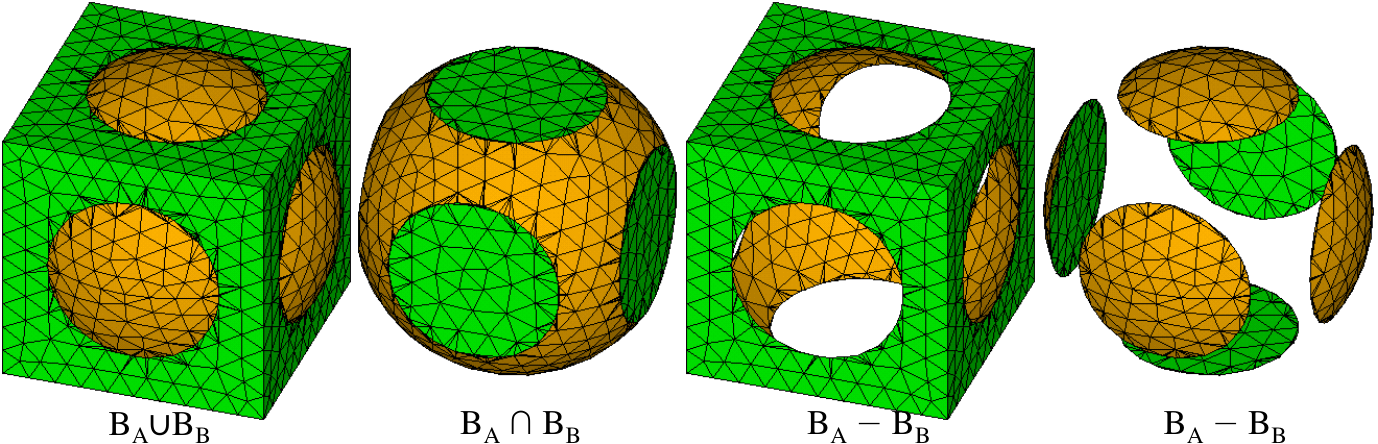} }
\caption{Boolean operations of a pair of blocks}
\label{fig:6:CubeSpere}       
\end{figure}

According to Algorithm \ref{alg:4:CreateSubBlock}, we can obtain all possibly produced sub-blocks. Now, there is a question that given two blocks B$_A$ and B$_B$ (Fig.\ref{fig:6a}), and after creating all sub-blocks, how we distinguish the union, intersection and subtraction. We develop a novel and simple method to solve this problem.

\newpage
\textbf{Step 1:} Obtain the non-subtraction(union and intersection) according to the orientation of closed intersection loops when assemble sub-blocks

In this step, we can only determine whether the sub-blocks are subtraction or not, and cannot distinguish they are the union or intersection clearly.

\begin{figure}[H]
\centering
\includegraphics[height = 5.1cm]{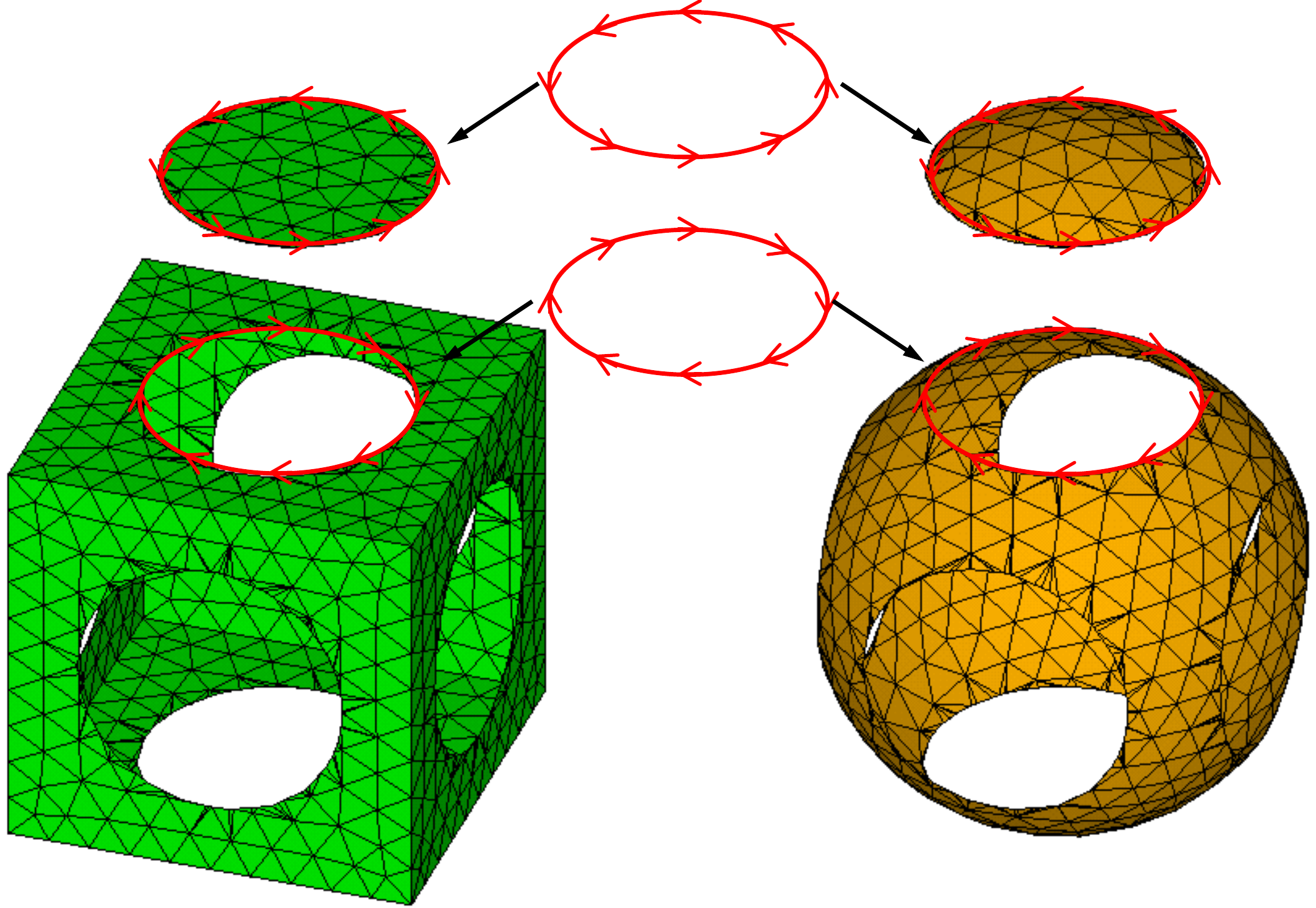}
\caption{Opposite loops and their owned or shared sub-surfaces}
\label{fig:7:Opposite}       
\end{figure}

For a sub-block SB, assuming it totally has $n$ closed loops, noted as $L_i$ ($i = 0,\ldots,n-1$). For each closed loop $L_i$, it has two opposite versions in which $L_i^+$ represents this loop with its original orientation and $L_i^-$ is the loop with same vertices as $L_i^+$ but opposite orientation (Fig.\ref{fig:7:Opposite}). These two loops are defined as \textit{twins} in Sect.\ref{sec:3}. The loop $L_i$ owns or shares two sub-surfaces denoted as SS$_{L_i}^A$ and SS$_{L_i}^B$ which come from two different original surfaces S$_A$ and S$_B$, respectively. we can easily determine the sub-block SB according to following conditions:

Case 1: if SS$_{L_i}^A$  and SS$_{L_i}^B$ are owned or shared by $L_i^+$ and $L_i^-$, respectively,
or if SS$_{L_i}^A$ and SS$_{L_i}^B$ are owned or shared by $L_i^-$ and $L_i^+$, respectively,
then the sub-block SB is union or intersection volume.

Case 2: if both SS$_{L_i}^A$ and SS$_{L_i}^B$ are owned or shared by $L_i^+$,
or if both SS$_{L_i}^A$ and SS$_{L_i}^B$ are owned or shared by $L_i^-$,
then the sub-block SB is subtraction volume.

\textbf{Step 2:} Distinguish the union firstly and then the intersections

As mentioned above, we let (B$_A$$\cup$B$_B$), (B$_A$$\cap$B$_B$), (B$_A$$-$B$_B$) and (B$_B$$-$B$_A$) represent the union, intersection, and subtractions of a pair of blocks B$_A$ and B$_B$. Obviously, we can receive the following relationship: (B$_A$$\cap$B$_B$) $\leq$ (B$_A$$\cup$B$_B$). The equivalent case holds only when B$_A$ and B$_B$ coincides, which has to be recognized and handled previously and thus is unnecessary to be considered at this time. Hence, we can in further has (B$_A$$\cap$B$_B$) $<$ (B$_A$$\cup$B$_B$). This means the intersection is always smaller than its corresponding union. This also means the minimum coordinates of all vertices of the intersection never smaller than those of the union while the maximum ones of the intersection never larger than those of the union. Therefore, the minimum and maximum coordinates of the vertices from both the union and intersection is exactly equal those only from the union.

Thus, we can distinguish the union and the intersection via following three sub-procedure.

Step 2-1: Obtain the maximum and minimum coordinates of both intersection and union; this can be done very easily by sorting all the vertices. In fact, this has been realized when merge and update the vertices. We can store their indexes there and call them here.

Step 2-2: Check each candidate union sub-block by comparing its maximum and minimum coordinates. If all of them are equivalent to their corresponding values obtained in above Step 2.1, then it must be the only union volume (Fig.\ref{fig:6c}).

Step 2-3: Set the rest sub-block(s) as the intersection volume (Fig.\ref{fig:6c}).

The above procedures are invalid only in the case that mesh B$_A$ is in B$_B$ or B$_B$ is in B$_A$. Hence, these special cases must be handled in pre-process.

\textbf{Step 3:} Determine all subtractions

After classifying the union and intersection, the subtractions (B$_A$$-$B$_B$) and (B$_B$$-$B$_A$) can be easily determined: define all sub-surfaces which compose the only union volume as \textit{outer} ones, while those from the intersection volume(s) as \textit{inner}, for each undistinguished sub-block,

(1) If one of the sub-surfaces from the sub-block is outer, and if the sub-surface belongs the original block B$_A$, then the sub-block is part of (B$_A$$-$B$_B$); else the sub-block is part of (B$_B$$-$B$_A$).

(2) If one of the sub-surfaces from the sub-block is inner, and if the sub-surface belongs the original block B$_A$, then the sub-block is part of (B$_B$$-$B$_A$); else the sub-block is part of (B$_A$$-$B$_B$).

After distinguishing, all triangles in inner sub-surfaces of subtractions have to reverse to obtain a valid topology of whole mesh model and make all the normal of facets outwards.

\section{Tests and results}
\label{sec:5}
\begin{figure}[H]
\centering
\subfigure[Original open surfaces]{
  \label{fig:8a}       
  \includegraphics[height=5cm]{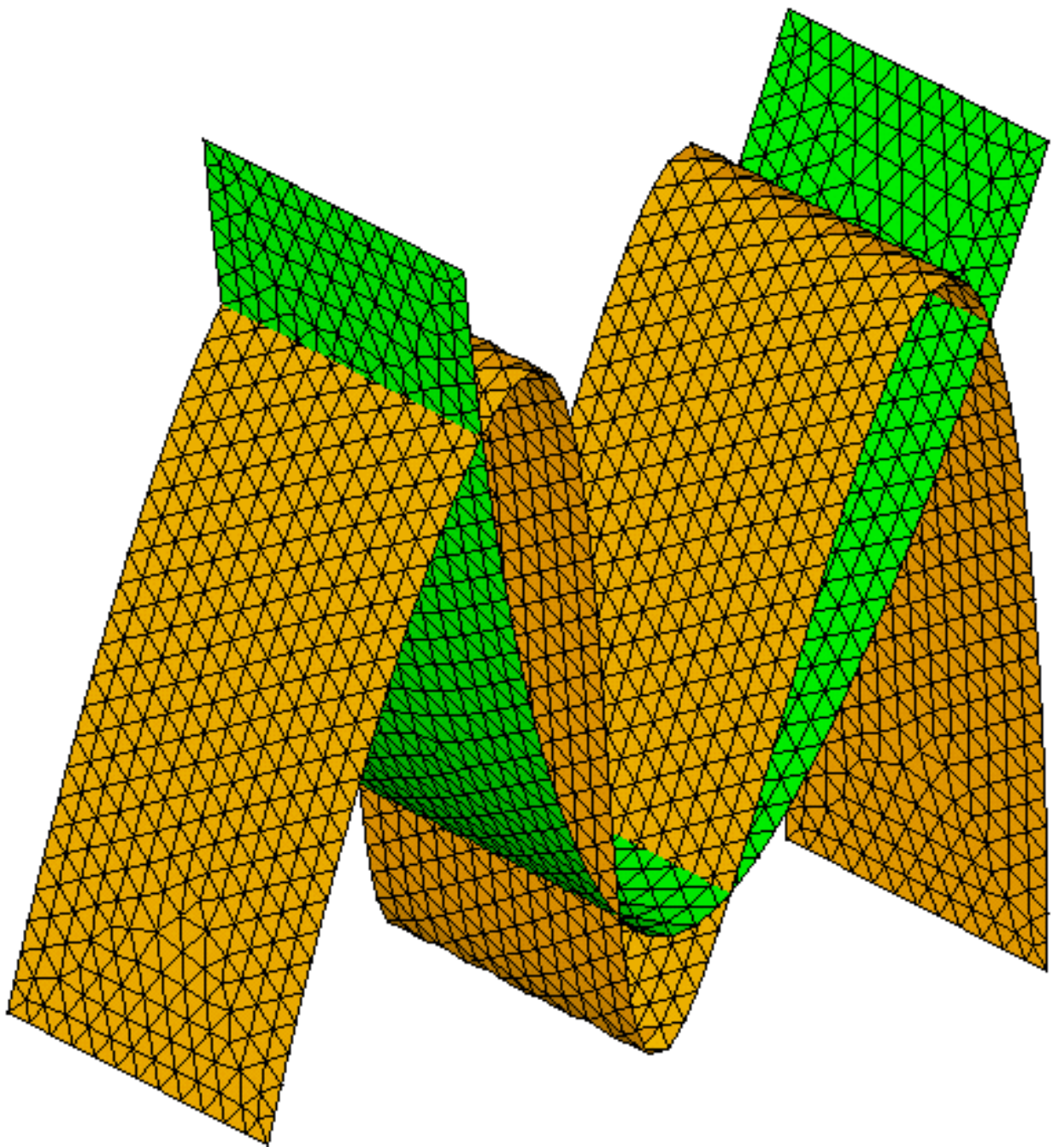} }
\hspace{0cm}
\subfigure[Sub-surfaces]{
  \label{fig:8b}       
  \includegraphics[height=5cm]{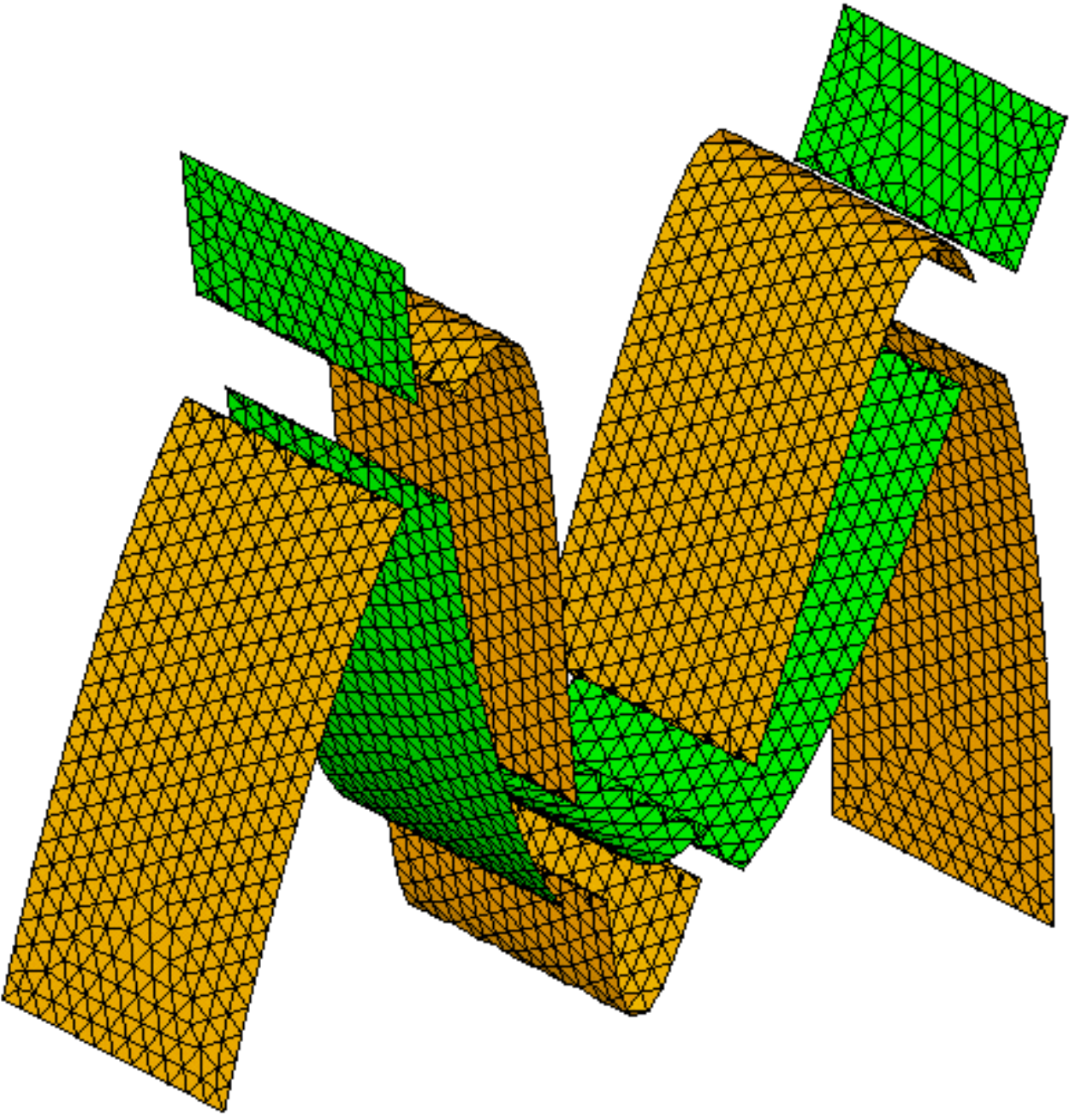} }
\caption{Intersection of a pair of open surfaces}
\label{fig:8:VandW}       
\end{figure}

\begin{figure}[H]
\centering
  \includegraphics[height=4.21cm]{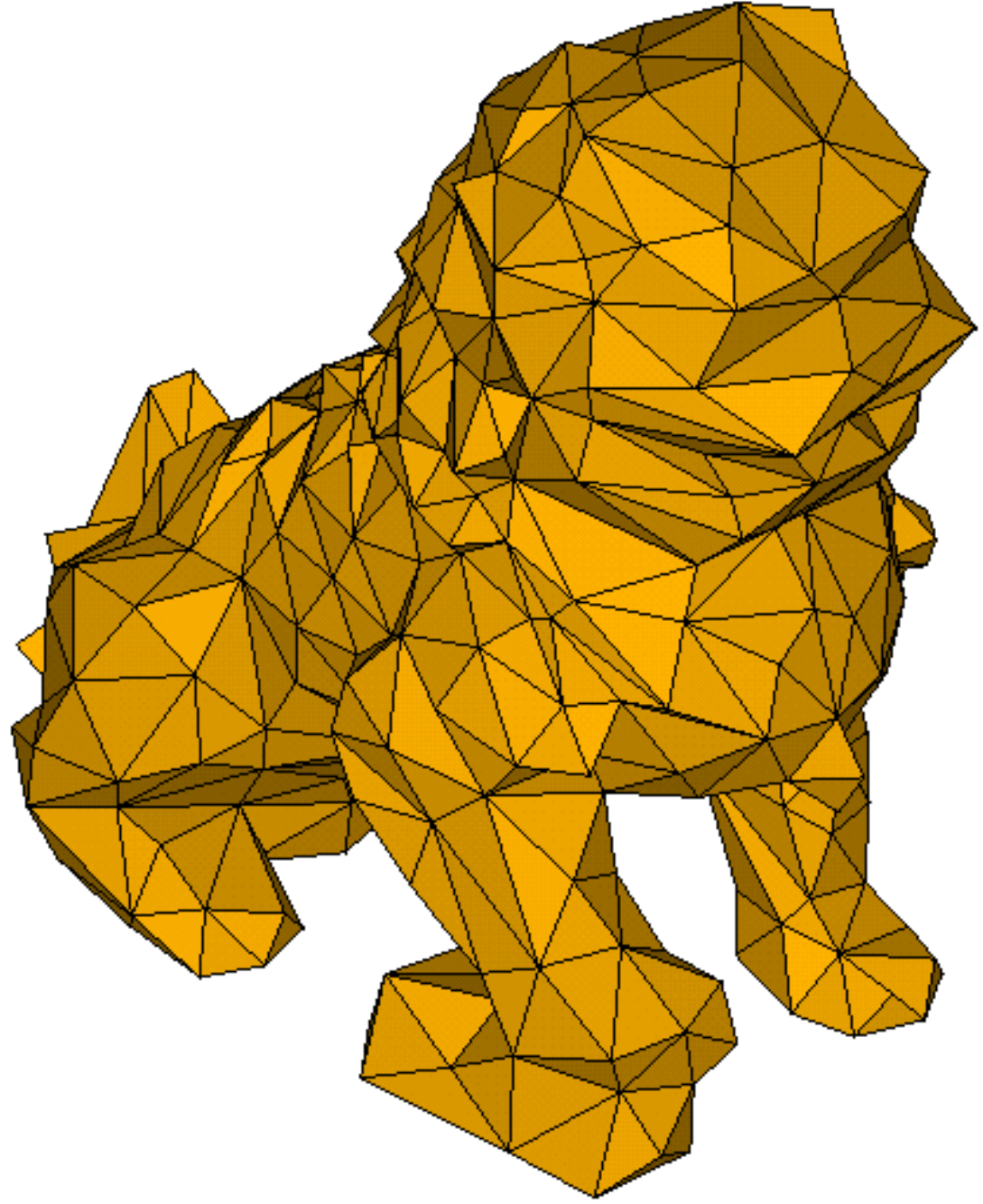}
\hspace{0cm}
  \includegraphics[height=4.21cm]{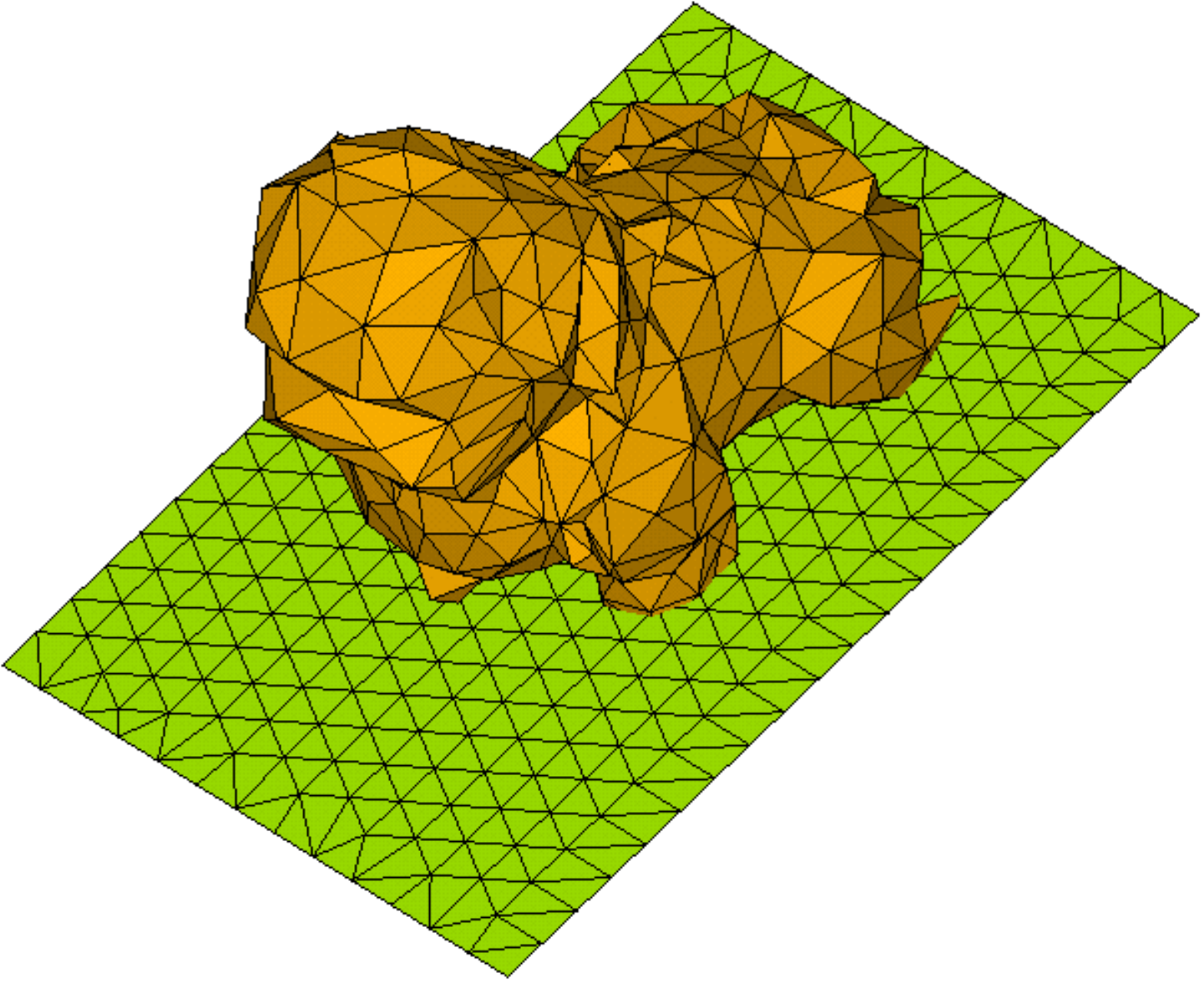}
\hspace{0cm}
  \includegraphics[height=4.21cm]{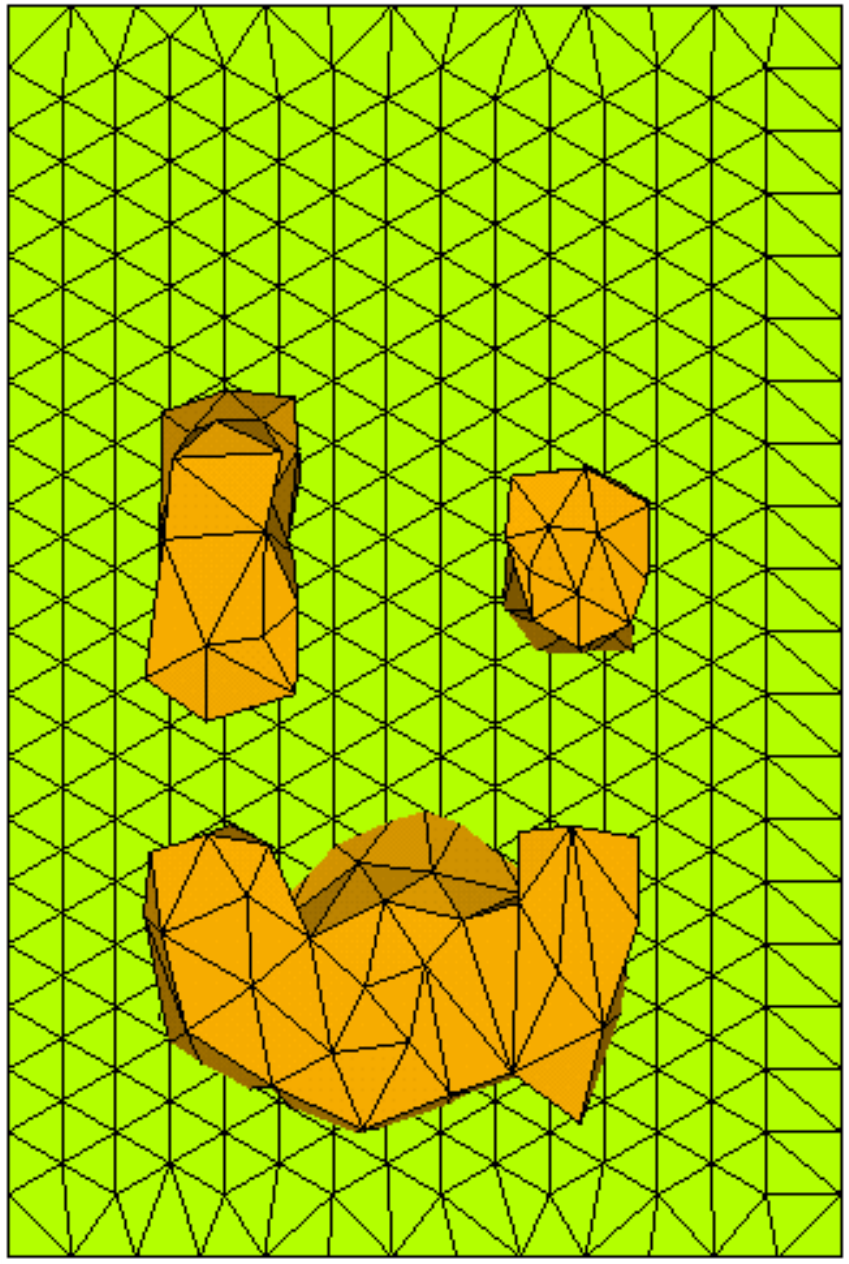}
\caption{A planar mesh and Chinese lion before dividing}
\label{fig:10}       
\end{figure}

\begin{figure}[H]
\centering
  \includegraphics[height=4.21cm]{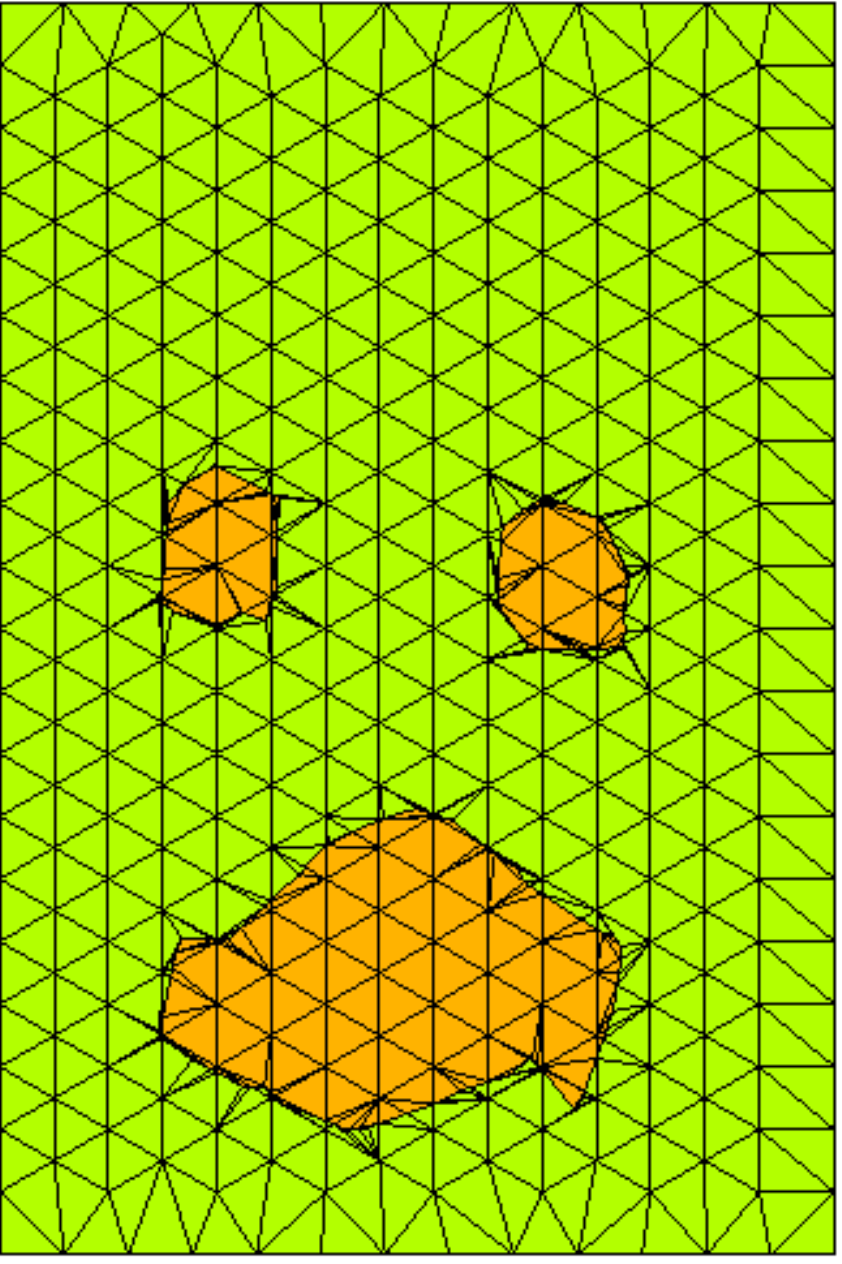}
\hspace{0cm}
  \includegraphics[height=4.21cm]{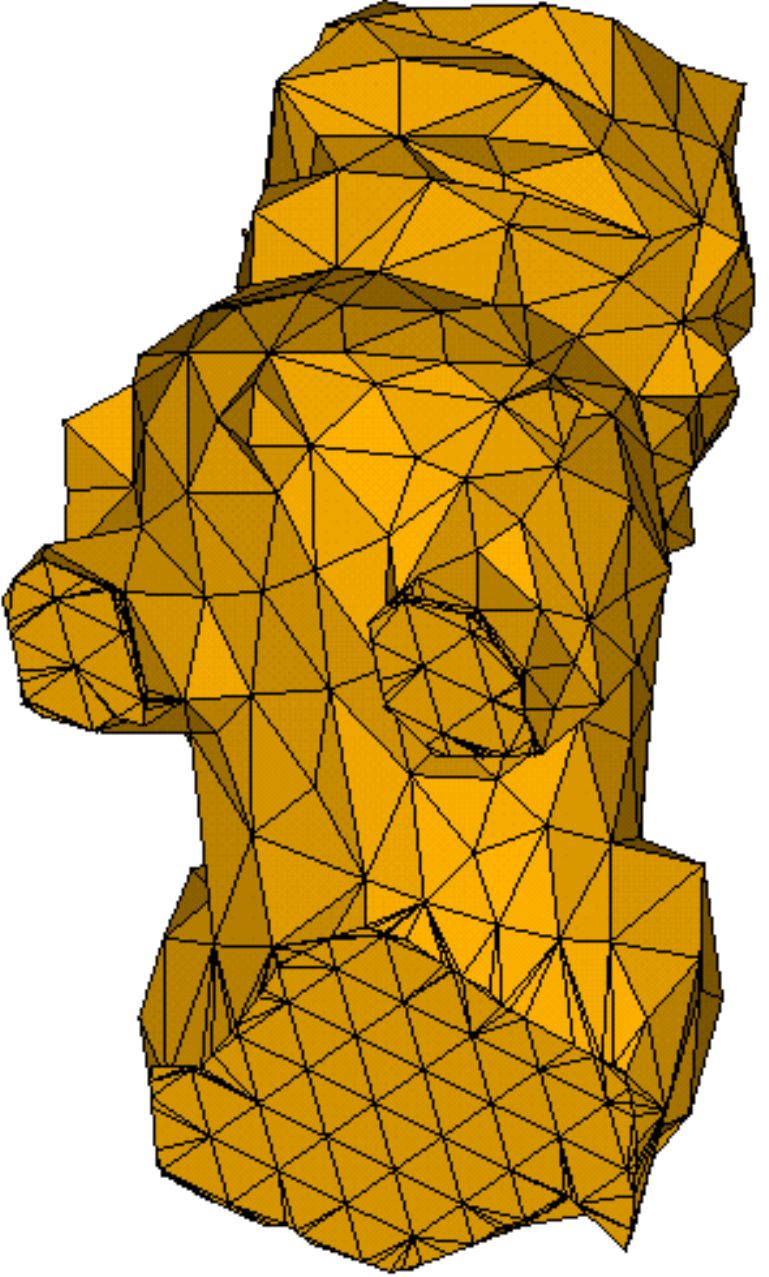}
\hspace{0cm}
  \includegraphics[height=4.21cm]{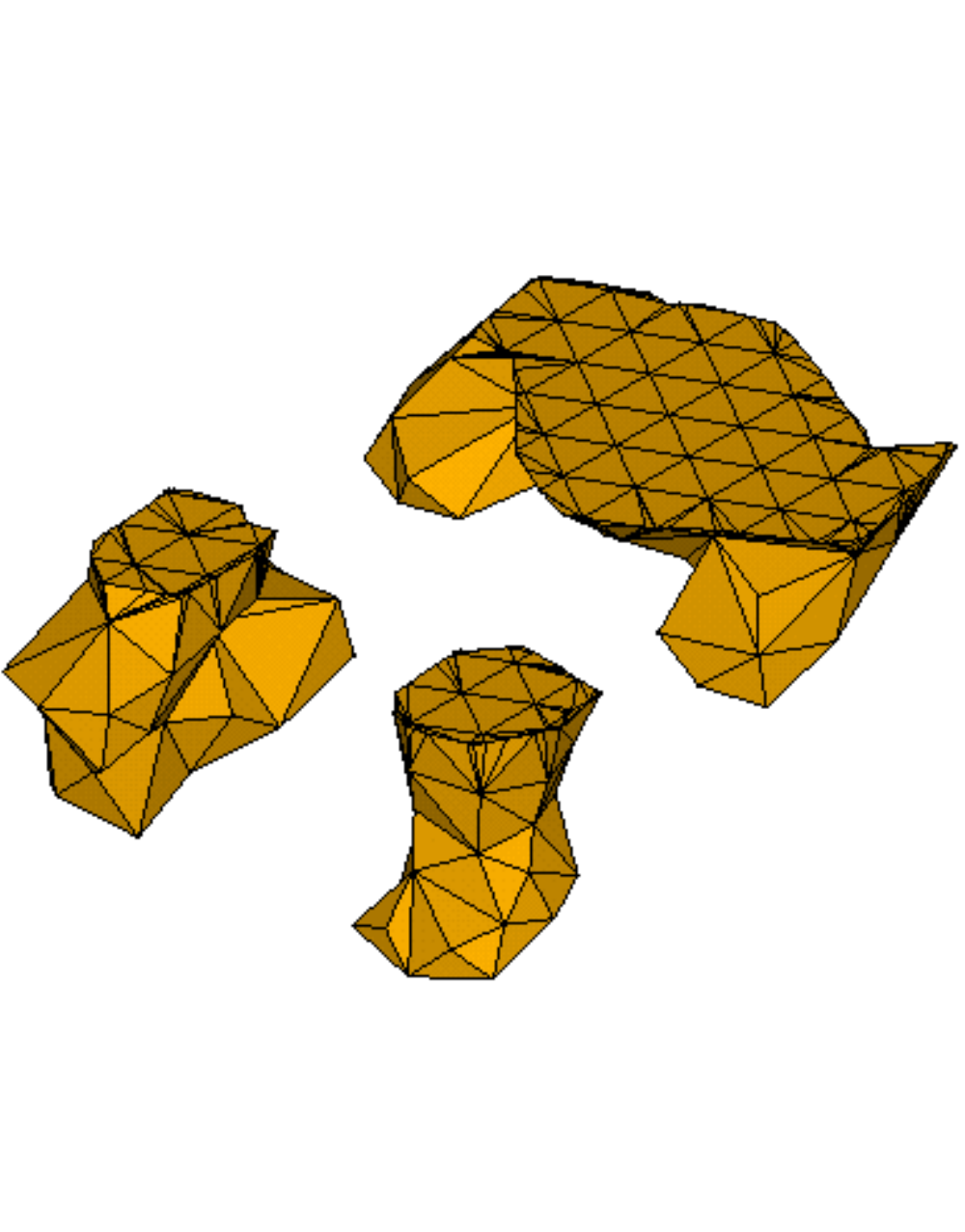}
\caption{A planar mesh and Chinese lion after dividing}
\label{fig:11}       
\end{figure}

\begin{figure}[H]
\centering
\subfigure[Original cylinders]{
  \label{fig:12a}       
  \includegraphics[height=4.5cm]{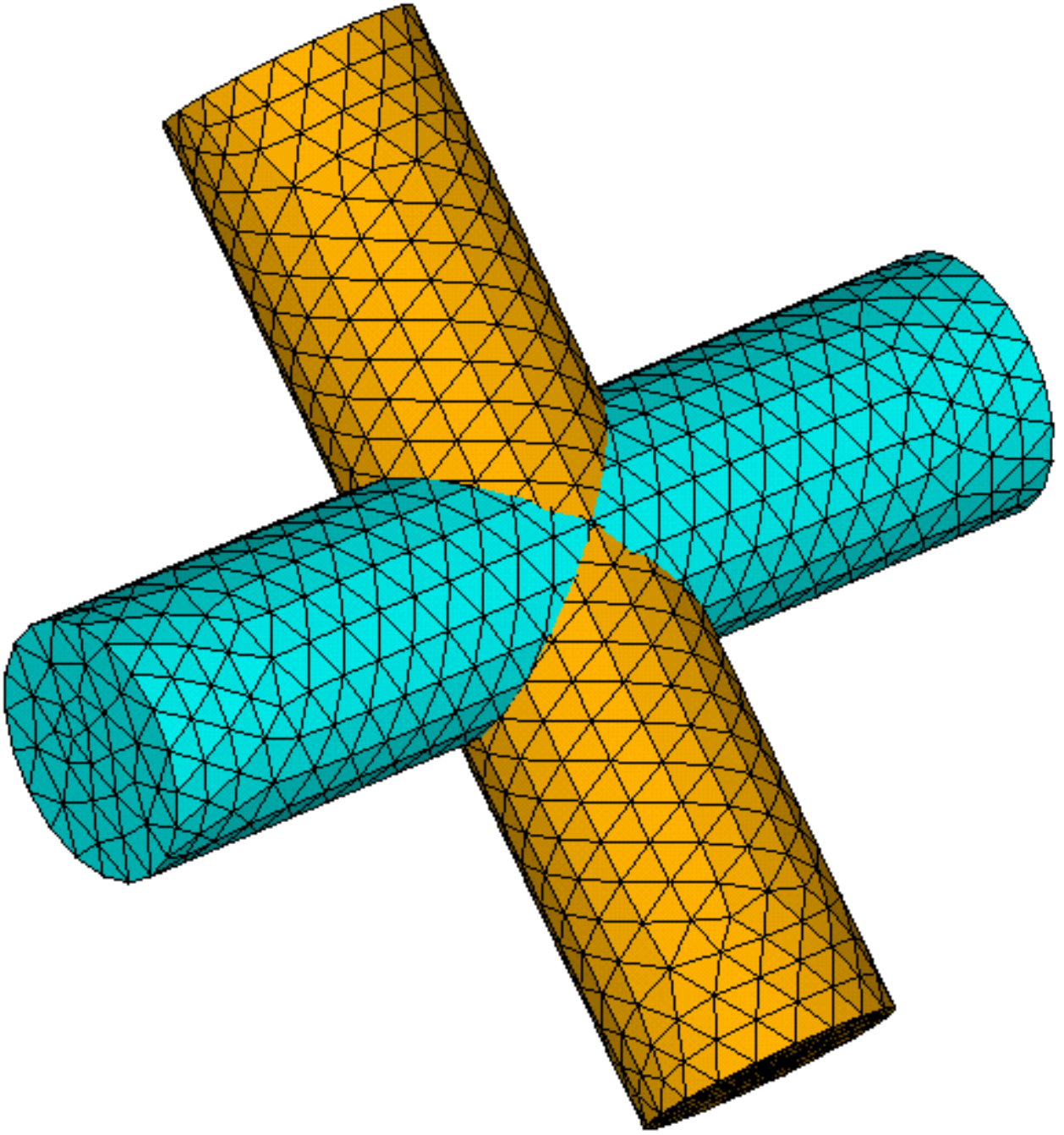} }
\hspace{0cm}
\subfigure[Sub-surfaces of a cylinder]{
  \label{fig:12b}       
  \includegraphics[height=4.5cm]{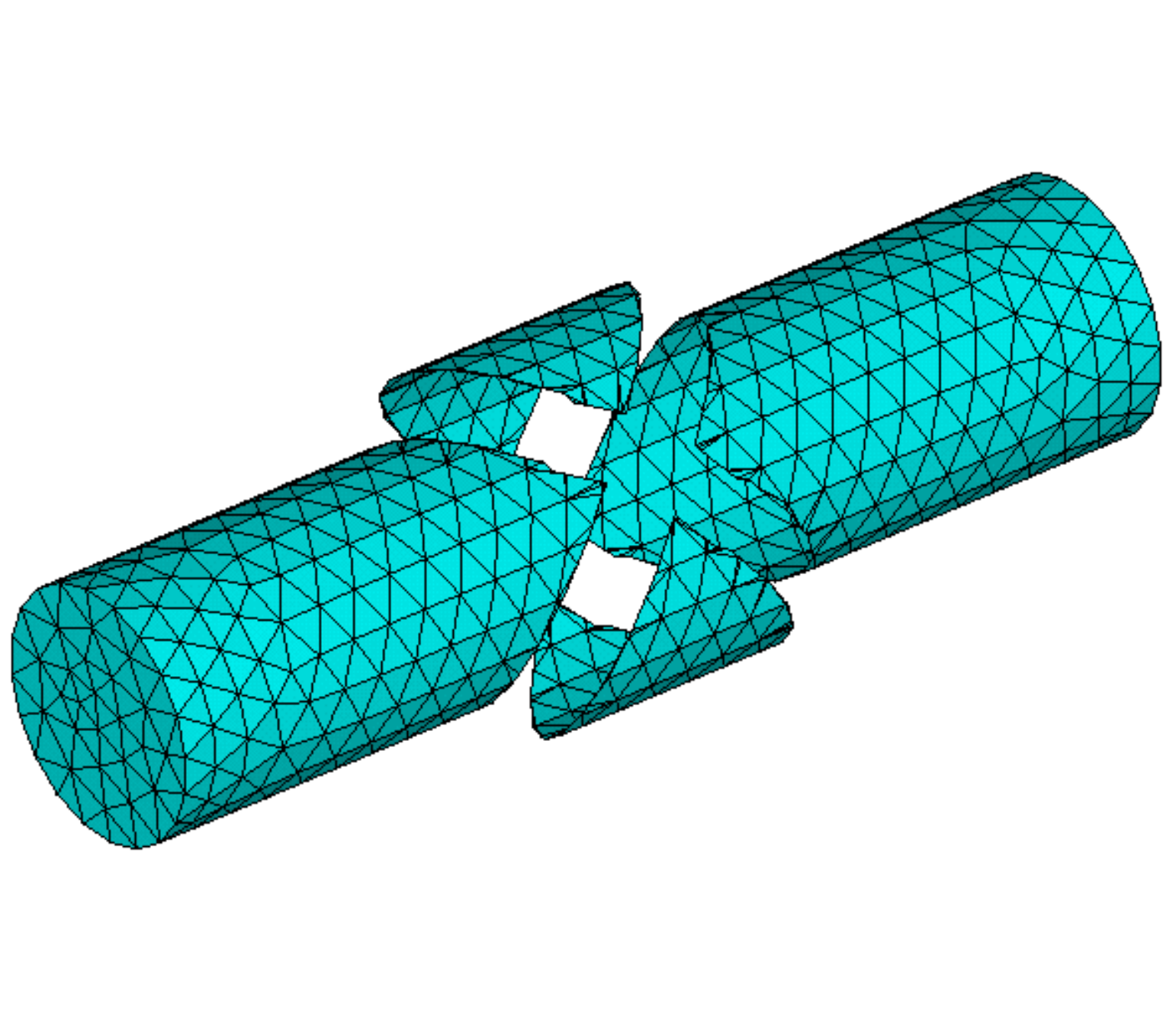} }
\hspace{0cm}
\subfigure[Intersection]{
  \label{fig:12c}       
  \includegraphics[height=4.5cm]{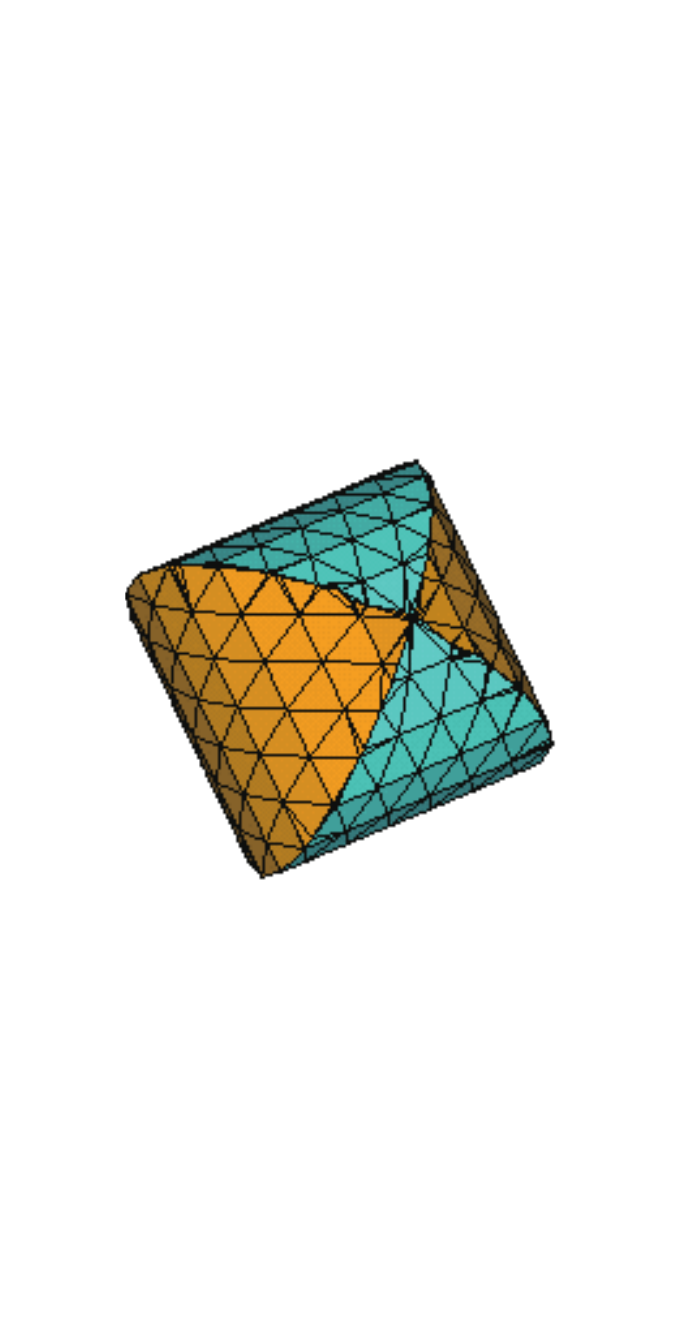} }
\hspace{0cm}
\subfigure[Union]{
  \label{fig:12d}       
  \includegraphics[height=4.5cm]{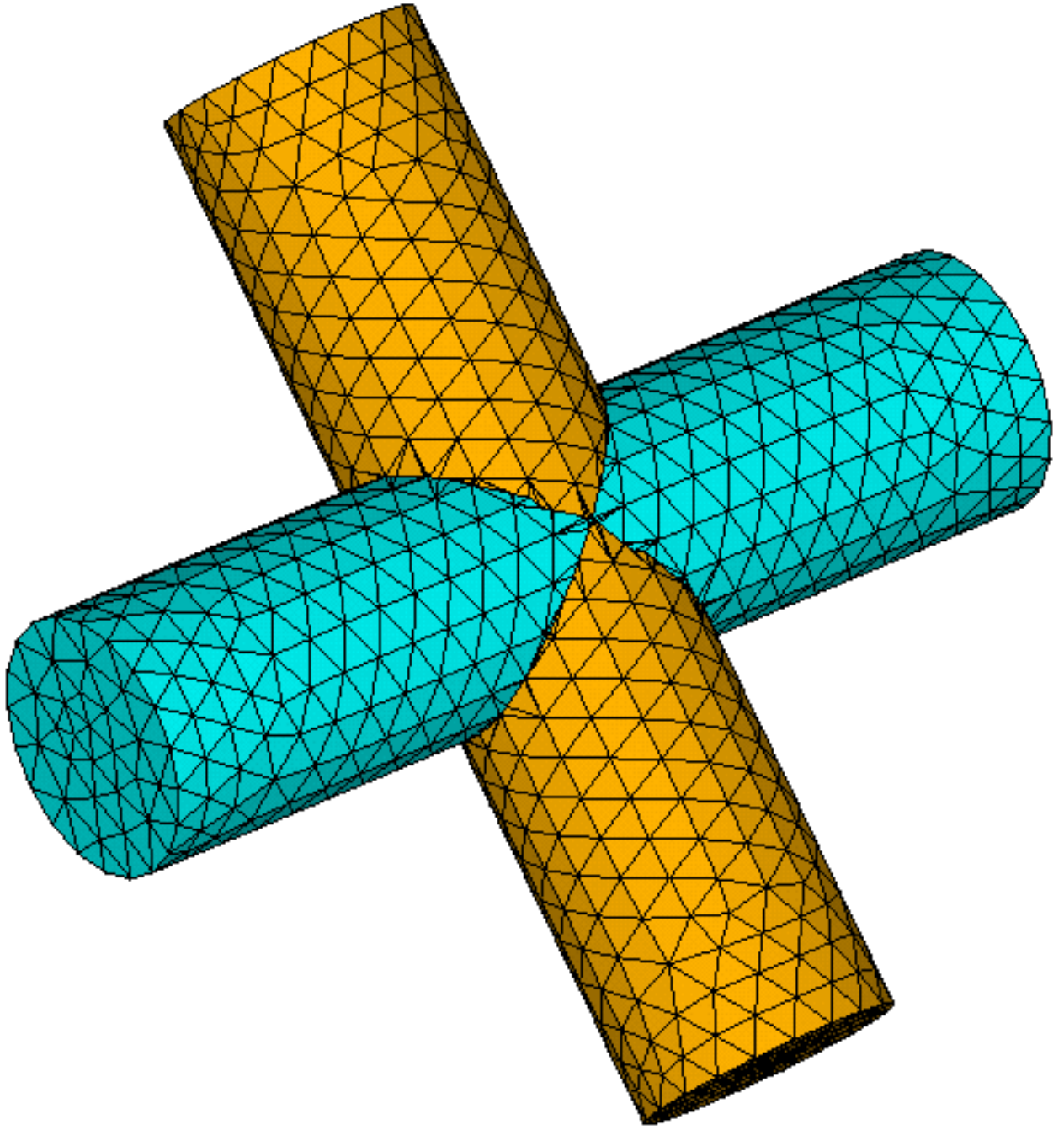} }
\hspace{0cm}
\subfigure[Subtractions]{
  \label{fig:12e}       
  \includegraphics[height=4.5cm]{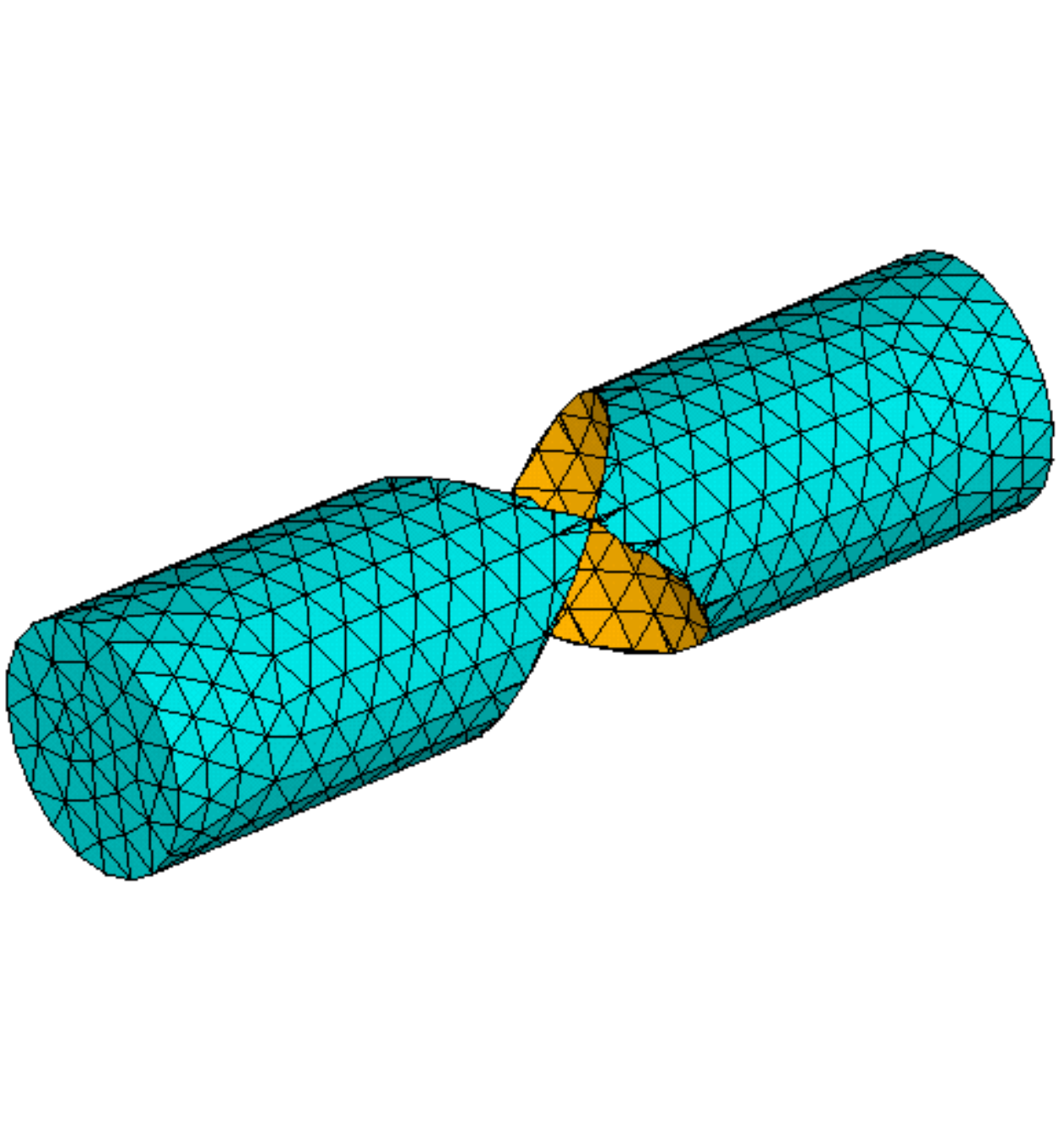}
  \hspace{0cm}
  \includegraphics[height=4.5cm]{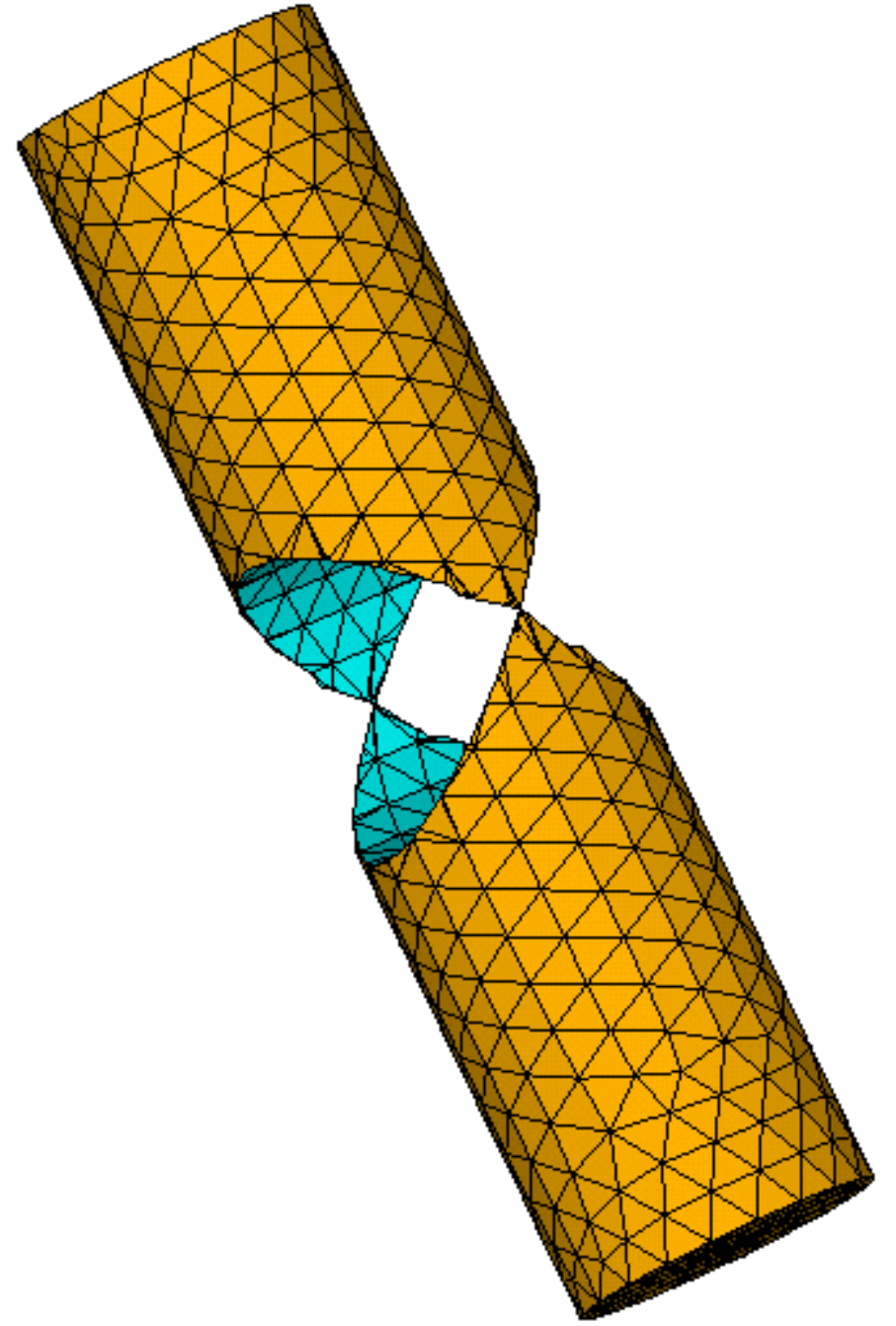}}
\caption{Boolean operations of a pair of cylinders}
\label{fig:12:Cylinder}       
\end{figure}

\begin{figure}[H]
\centering
\subfigure[Original pair of torus]{
  \label{fig:13a}       
  \includegraphics[height=4.5cm]{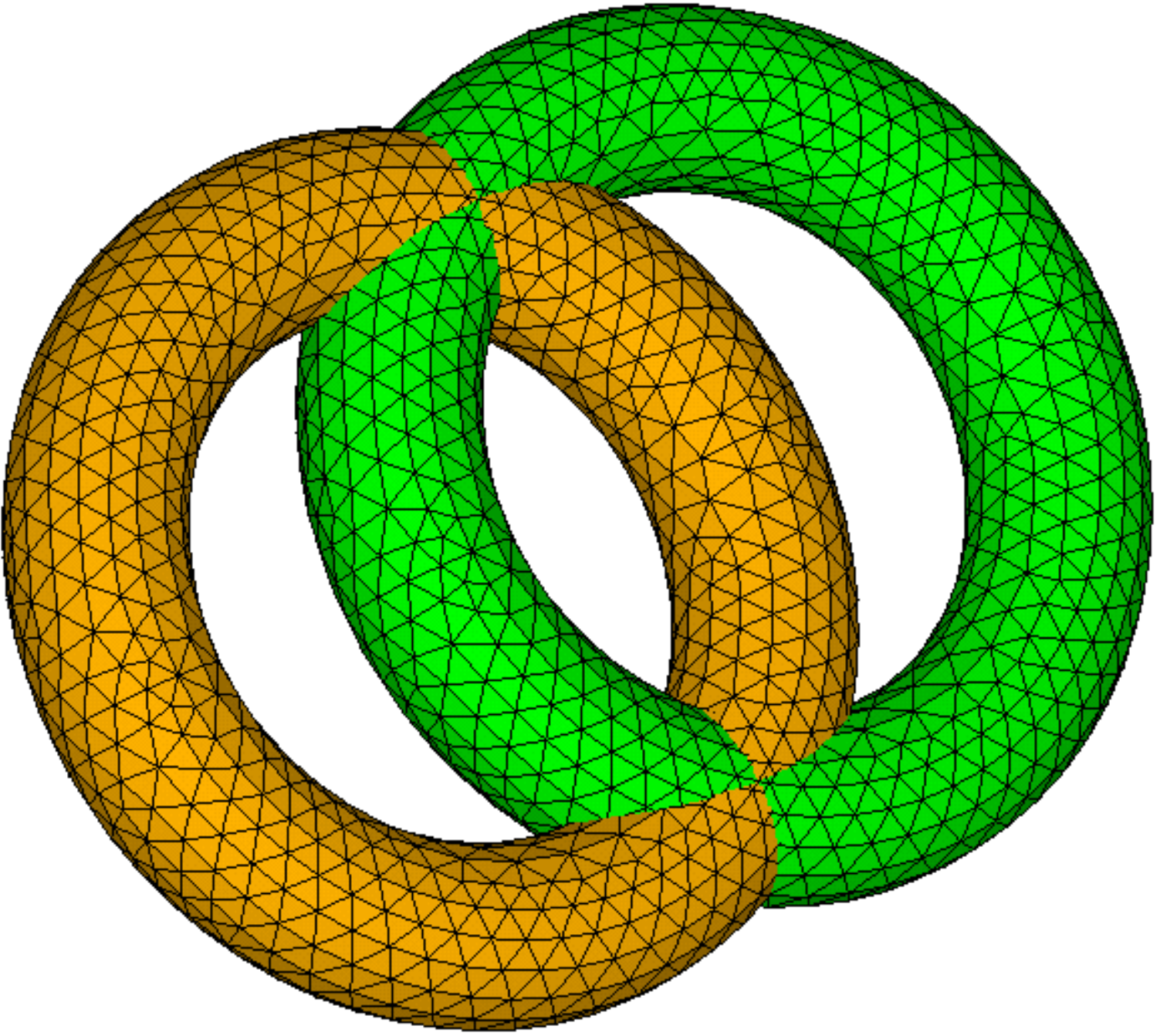} }
\hspace{0cm}
\subfigure[Union]{
  \label{fig:13b}       
  \includegraphics[height=4.5cm]{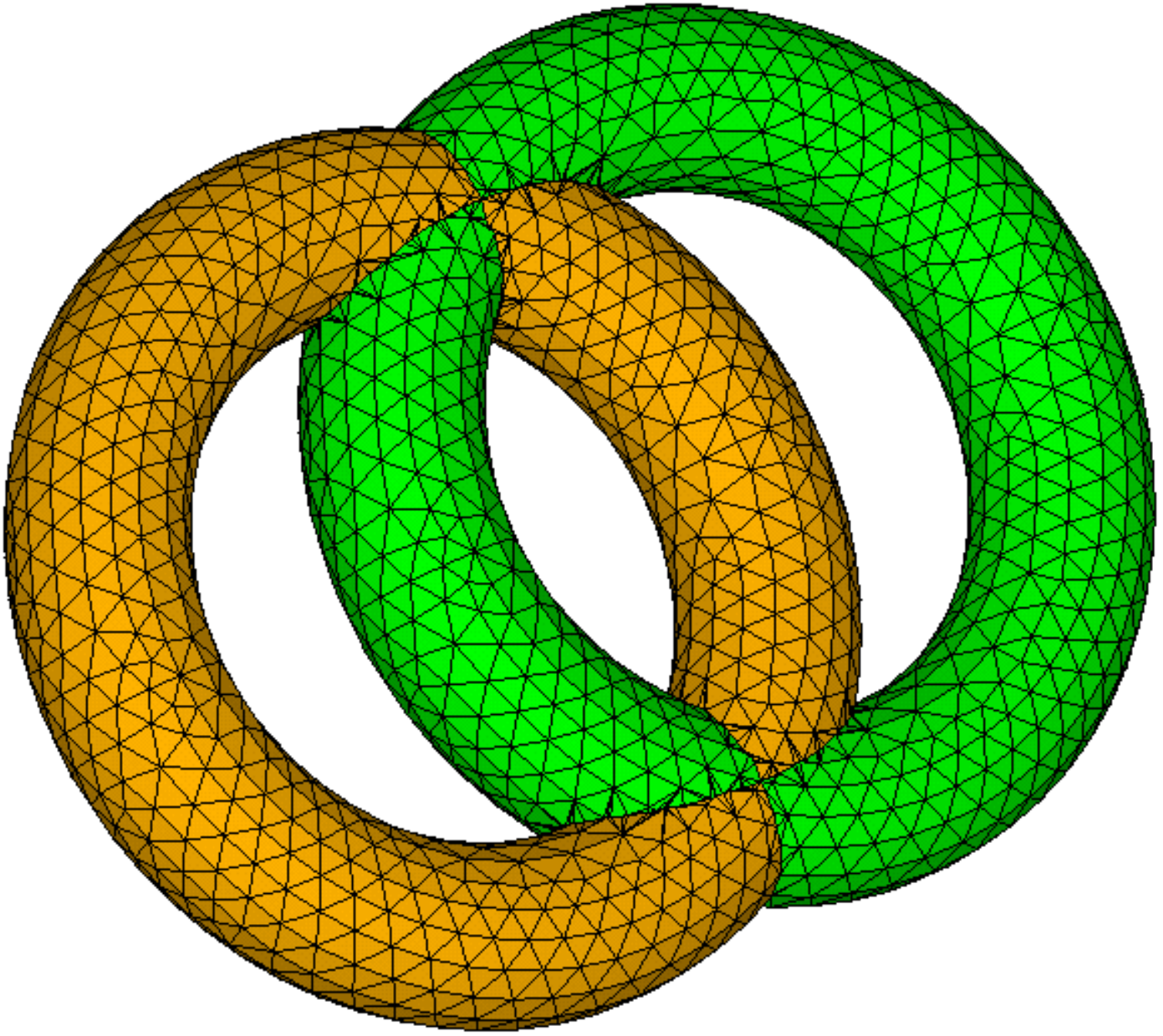} }
\hspace{0cm}
\subfigure[Intersection]{
  \label{fig:13c}       
  \includegraphics[height=4.5cm]{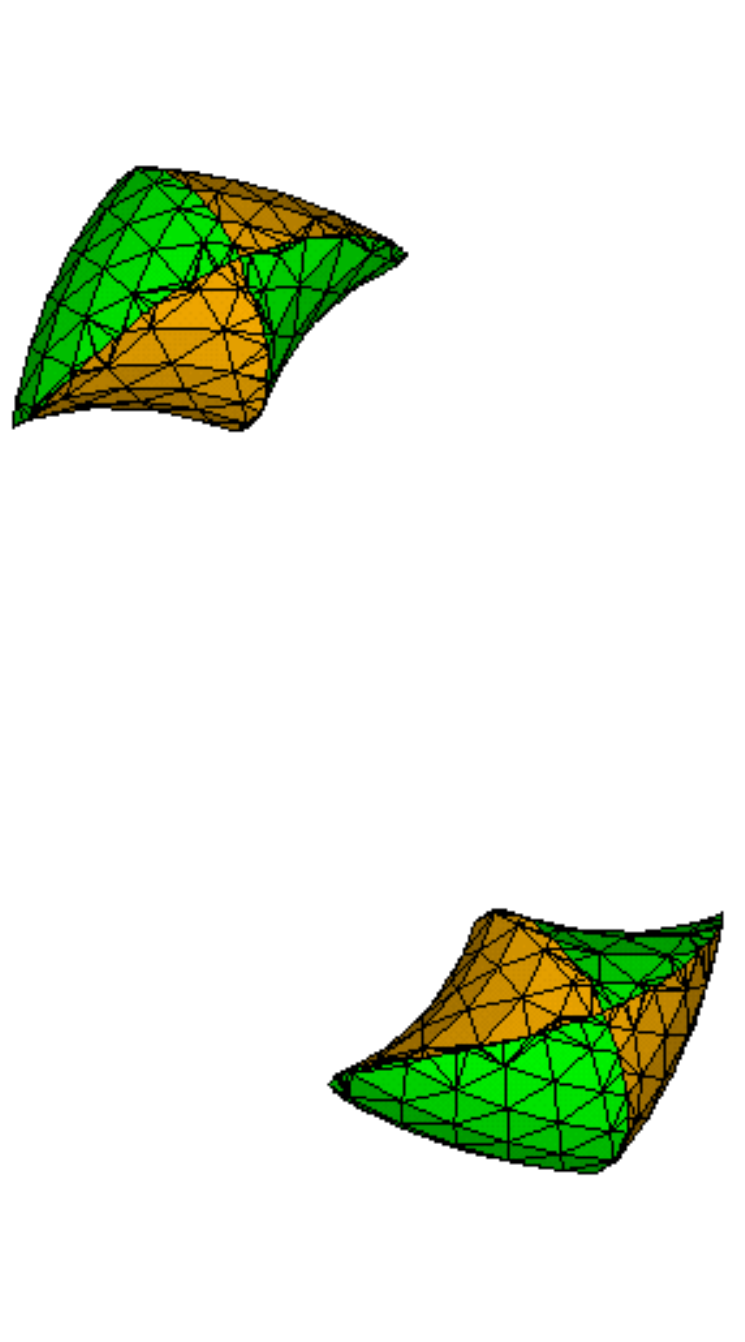} }
\hspace{0cm}
\subfigure[Subtractions]{
  \label{fig:13d}       
  \includegraphics[height=4.5cm]{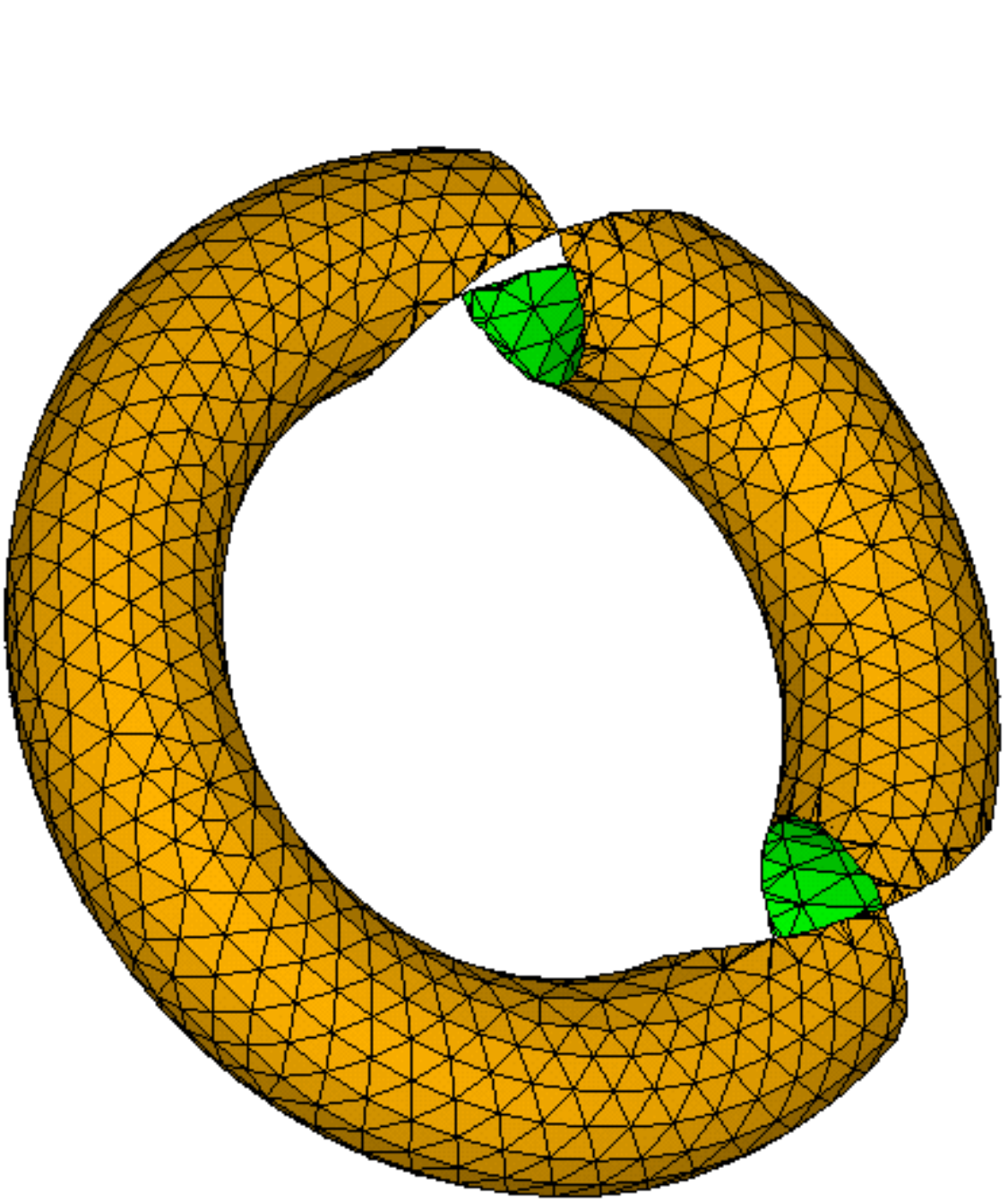}
  \hspace{0cm}
  \includegraphics[height=4.5cm]{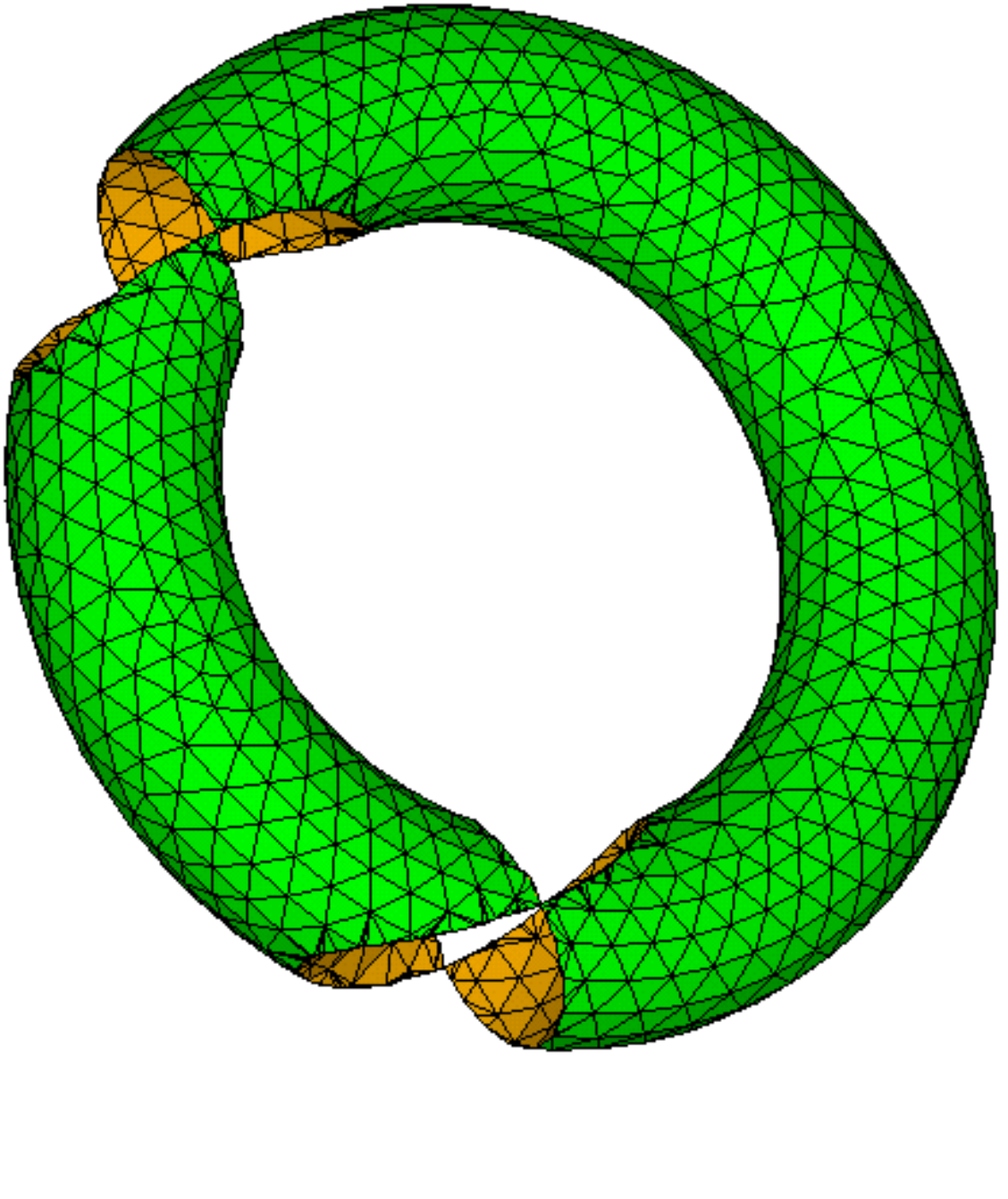}}
\caption{Boolean operations of a pair of torus}
\label{fig:13:Torus}       
\end{figure}

As described in Sect.\ref{sec:1}, we aim at performing Boolean operations only over a pair of surfaces. Either of the surfaces can be open or closed. In this section, we obtain the Boolean operations of several pairs of surfaces including open-and-open (Fig.\ref{fig:8:VandW}), open-and-closed (Fig.\ref{fig:10} and Fig.\ref{fig:11}) and closed-and-closed (Fig.\ref{fig:12:Cylinder} and Fig.\ref{fig:13:Torus}) to demonstrate our approach.

The intersection of a pair of open triangulated surfaces is tested in Fig.\ref{fig:8:VandW}. Before considering the boundary outer loops of the original surfaces, four open intersection loops can be formed for each surface; and then the boundary loop is divided into five closed loops; hence five corresponding sub-surfaces can be created for each surface.

The division of a planar meshed surface with a triangulated Chinese lion is presented to test the Booleans for a open surface with a closed surface. The model Chinese lion (Fig.\ref{fig:10}) is obtained from http://shapes.aim-at-shape.net/. From Fig.\ref{fig:10}, we can see that three closed loops can be formed in both the Chinese lion and the surface. Figure.9 displays the dividing results that four sub-surfaces can be created for both of original surfaces, and the lion is divided into four sub-blocks.

To test Booleans on two closed surfaces, we give the examples of Boolean operations for a pair of cylinders (Fig.\ref{fig:12:Cylinder}) and a pair of torus (Fig.\ref{fig:13:Torus}). After computing the intersection edges of triangles for cylinders, four \textit{soft} closed (defined in Sect.\ref{sec:4.4}) intersection loops can be formed. Both the first vertex and the last vertex of each soft closed loop is shared by four intersection edges. Based on these soft closed loops, four sub-surfaces can be created for each cylinder, and then sub-blocks such as union, intersection and subtractions can be easily assembled and distinguished according to the sub-faces. Figure.\ref{fig:13:Torus} shows the Boolean operations for a pair of torus which have the same inner radius. It is similar to that of cylinders but a little more complex. The above two examples only produced soft closed loops, a simple example in which hard closed loops are obtained is presented in Fig.\ref{fig:6:CubeSpere}.

\section{Conclusions}
\label{sec:6}

We present a simple and robust approach to perform Boolean operations on a pair of manifold triangulated surfaces. Our approach falls into the type of exact arithmetic methods in the aspect of computation. We use Octree to locate and find potentially intersected triangles and adopt parallel algorithm to calculate the intersection lines of triangles. We form and define the intersection loops into \textit{open}, \textit{hard closed} and \textit{soft closed} ones, and then create sub-surfaces by growing while only the closed intersection loops are set as boundary loops of sub-surfaces and deemed as the initial advancing fronts. Sub-blocks are assembled very easily according to boundary loops of sub-surfaces, and then distinguished by comparing min and max coordinates of updated vertices.

There are two stages in our approach: the first, which is to calculate intersection lines of triangles and then merge and update the resulting surfaces, needs coordinates calculations; the second, which includes forming intersection loops, creating sub-surfaces, assembling and distinguishing sub-blocks, is only based on the cleared and updated topology of triangular meshes while without coordinate computations of geometric entities. Effectiveness of our approach has been illustrated by several test examples.

%
%

%
%



\end{document}